\documentclass[aps,superscriptaddress]{revtex4}
\usepackage{amssymb,amsmath,epsfig}
\usepackage[colorlinks=true, pdfstartview=FitV, linkcolor=blue, citecolor=red, urlcolor=magenta, breaklinks=true]{hyperref}
\usepackage{graphicx}
\usepackage{epstopdf}

\usepackage{subfig}

\begin{document}
\title{Diffusive process under Lifshitz scaling and pandemic scenarios}

\author{M.A. Anacleto}\email{anacleto@df.ufcg.edu.br}
\affiliation{Unidade Acad\^{e}mica de F\'{\i}sica, Universidade Federal de Campina Grande
Caixa Postal 10071, 58429-900 Campina Grande, Para\'{\i}ba, Brazil}

\author{F.A. Brito}\email{fabrito@df.ufcg.edu.br}
\affiliation{Unidade Acad\^{e}mica de F\'{\i}sica, Universidade Federal de Campina Grande
Caixa Postal 10071, 58429-900 Campina Grande, Para\'{\i}ba, Brazil}
\affiliation{Departamento de F\'isica, Universidade Federal da Para\'iba, 
Caixa Postal 5008, 58051-970 Jo\~ao Pessoa, Para\'iba, Brazil}

\author{A. R. de Queiroz}\email{amilcarq@gmail.com}
\affiliation{Unidade Acad\^{e}mica de F\'{\i}sica, Universidade Federal de Campina Grande
Caixa Postal 10071, 58429-900 Campina Grande, Para\'{\i}ba, Brazil}

\author{E. Passos}\email{passos@df.ufcg.edu.br}
\affiliation{Unidade Acad\^{e}mica de F\'{\i}sica, Universidade Federal de Campina Grande
Caixa Postal 10071, 58429-900 Campina Grande, Para\'{\i}ba, Brazil}

\author{J.R.L. Santos}\email{joaorafael@df.ufcg.edu.br}
\affiliation{Unidade Acad\^{e}mica de F\'{\i}sica, Universidade Federal de Campina Grande
Caixa Postal 10071, 58429-900 Campina Grande, Para\'{\i}ba, Brazil}

\begin{abstract}

We here propose to model active and cumulative cases data from COVID-19 by a continuous effective model based on a modified diffusion equation under Lifshitz scaling with a dynamic diffusion coefficient. The proposed model is rich enough to capture different aspects of a complex virus diffusion as humanity has been recently facing. The model being continuous it is bound to be solved analytically and/or numerically. So, we investigate two possible models where the diffusion coefficient associated with possible types of contamination are captured by some specific profiles.  The active cases curves here derived were able to successfully describe the pandemic behavior of Germany and Spain. Moreover, we also predict some scenarios for the evolution of COVID-19 in Brazil. Furthermore, we depicted the cumulative cases curves of COVID-19, reproducing the spreading of the pandemic between the cities of S\~ao Paulo and S\~ao Jos\'e dos Campos, Brazil. The scenarios also unveil how the lockdown measures can flatten the contamination curves. We can find the best profile of the diffusion coefficient that better fit the real data of pandemic.
 
\end{abstract}

\maketitle
\pretolerance10000

\section{Introduction}
\label{intro}

In December 2019, the world started to face a new type of severe pneumonia which appeared in Wuhan, China. Only two months later the International Committee on Taxonomy of Viruses named the virus responsible for these pneumonia cases as severe acute respiratory syndrome coronavirus 2 or SARS-CoV-2, whose disease was popularly known as coronavirus disease 2019, or simply COVID-19 \cite{yang_20}. Such disease was classified as a public health emergency of international concern at the end of January 2020, by the World Health Organization. Up to now, SARS-CoV-2 has spread all over the world, presenting more than $3.5$ million of cases, taken approximately $247107$ lives, and has a new epicenter in the United States of America which has reported almost one-third of the total amount of cases \cite{jh_covid}. Few places in the world were able to fully control the pandemic of COVID-19. In Europe, for instance, Italy and Spain were for a long time the world's two worst-hit countries by the COVID-19. Now, the numbers of active cases in these two countries are slowly decreasing and they are facing the final stage of the pandemic. One of the best countries in Europe to adopt measures against COVID-19 so far is Germany, who is also entering in the controllable phase of the pandemic. Behind the success of Germany are measures of social distance or lockdown procedures, and a large number of people tested for SARS-CoV-2 \cite{world}. 

After spreading in Asia and Europe, now COVID-19 is a challenge for the USA, as well as for low and middle-income countries, such as Brazil. There is a serious concern on the international scientific community about the behavior of SARS-CoV-2 in such countries, since they face other severe problems such as poverty, food security, economic growth, besides other diseases like human immunodeficiency virus, tuberculosis, and malaria \cite{imp_report_19}.
 
So far, several scientific works and reports based on numerical simulations have been published, lighting the evolution of the pandemic in different countries and reflecting the actions as well as the strategies of each country to mitigate the effects of COVID-19. Some of these studies can be found in references \cite{imp_report_19, imp_rep_0320, petra_20, li_20, marciano_20}.  Among recent works on this subject, we also highlight an interesting proposal of an age-structured model presented by Canabarro et al. \cite{canabarro_2020}, a model based on hospital infrastructure developed by Pacheco et al. \cite{pacheco_2020}, and a study to predict COVID-19 peaks around the world based on active cases curves, introduced by Tsallis et al. \cite{tsallis_2020}. In this work we intend to collaborate with the current investigations by proposing a model which can describe the evolution of the pandemic through the solutions of a modified version of the diffusion equation.   

The standard diffusion equation describes the macroscopic behavior due the effect of many micro-particle bodies, as it is observed in a Brownian motion, for instance \cite{brown}. An interesting modification of this equation was proposed in the seminal paper of Petr Horava \cite{Horava-PRL}, in his studies about quantum gravity, where he extended the definition of spectral dimension to theories on smooth spacetimes in anisotropic or Lifshitz scaling. In our investigation, we introduce a new version of the diffusion equation inspired by Horava's work, and we use it to fit real active cases data of COVID-19 from Germany, Spain and Brazil. The diffusion equation with the Lifshitz scalling and the equation of motion for the diffusion coefficient are going to be introduced in section  \ref{sec01}. In section \ref{sol} we are going to show different solutions for the diffusion equation which can be used to fit the evolution of COVID-19. The availability of our models are carefully discussed in section \ref{act_cases}. The spreading of the pandemic between two different cities is modeled in section \ref{coupled}. Then, we present our final remarks and perspectives in section \ref{conclusions}.

\section{Diffusion equation and Lifshitz scaling}
\label{sec01}
The first issue to describe a pandemic evolution consists in choosing an appropriate diffusion process. The complexity of a virus transmission such as SARS-CoV-2, demands a diffusion process characterized by a probability density $\rho=\rho({\bf x},\tau;{\bf x}',\tau';\sigma)$ measuring the diffusion from a time $\tau$ to a time $\tau'$, and from a space coordinate ${\bf x}$ to ${\bf x}'$ at a diffusion time $\sigma$. Notice that $\sigma$ and $\tau$ are two different types of time, $\tau$ would be understood as the standard time variation of the pandemic, while $\sigma$ would control the collective response to the pandemic, such as social distance measures, for instance. Besides, as each country can adopt several strategies to mitigate the pandemic effects, it is expected that the diffusion process would account for different degrees of anisotropy. A general continuous diffusion equation that attends such criteria was introduced by Horava in his seminal work \cite{Horava-PRL}, whose form is 
\begin{equation}
\frac{\partial}{\partial\sigma}\rho({\bf x},\tau;{\bf x}',\tau';\sigma)=\left(\frac{\partial^2}{\partial\tau^2} +(-1)^{z+1}\Delta^z\right)\,\rho({\bf x},\tau;{\bf x}',\tau';\sigma)\,.
\end{equation}
Here $\tau$ is the so-called Euclidean time and $z$ is the Lifshitz critical exponent, which measures the anisotropic scaling of a given model \cite{Horava-PRL,bp}. The Lifshitz critical exponent is essential to determine the spectral dimension, which can be applied to several geometric objects presenting fractal behavior \cite{Horava-PRL}. The relative sign $(-1)^{z+1}$ concerns the requirement of ellipticity of the diffusion operator valid for integer $z$, but the results can be analytically continued for any positive real $z$ \cite{Horava-PRL,bp}. An extra relevant ingredient to proper modeling a pandemic spread is a diffusion coefficient, which can account for the transmission rate of the virus. Therefore, this discussion suggests that a pandemic scenario is governed by the following anisotropic diffusion equation 
\begin{eqnarray}\label{eq1}
\frac{\partial}{\partial\sigma}\rho({\bf x},\tau;{\bf x}',\tau';\sigma)=\phi(\tau)\left(\frac{\partial^2}{\partial\tau^2} +(-1)^{z+1}\Delta^z\right)\phi(\tau)^{\,-1}\,\rho({\bf x},\tau;{\bf x}',\tau';\sigma),
\end{eqnarray}
where  $\phi(\tau)$ is a dynamic diffusion coefficient. We are going to show that a proper balance between $\phi$ and $z$ yields to distributions that can fit real pandemic data. Let us also constrain the diffusion coefficient $\phi(\tau)$ with the standard Lagrangian 
\begin{equation}
L=-\frac{1}{2}\,\phi_{\tau}^{\,2}-V(\phi)\,;\qquad \phi_\tau=\frac{d\,\phi}{d\,\tau}\,,
\end{equation}
where the negative kinetic part stands for the Euclidean time. Therefore, the equation of motion for the diffusion coefficient is such that
\begin{equation}
\phi_{\,\tau\tau}=V_{\,\phi}\,.
\end{equation}
By integrating the previous equation once, we find the first-order differential equation
\begin{equation}
\phi_{\,\tau}=\pm\,W_{\phi}\,,
\end{equation}
where we considered
\begin{equation}
V(\phi)\,\equiv \frac{W_{\,\phi}^2}{2}\,; \qquad W_{\,\phi}=\frac{d\,W}{d\,\phi}\,.
\end{equation}
The function $W$ is known as superpotential in analogy with the bosonic sector of a super symmetric field theory.

\section{Solutions}
\label{sol}

The diffusion equation (\ref{eq1}) has the general solution
\begin{eqnarray}\label{eq2}
\rho({\bf x},\tau;{\bf x}',\tau';\sigma)=\rho_0\,\int{\frac{d\omega\, d^D{\bf k}}{(2\pi)^{D+1}}\,\phi(\tau)\,e^{i\omega(\tau-\tau')+i{\bf k}\cdot({\bf x}-{\bf x}')}\,e^{-\sigma(\omega^2+|{\bf k}|^{2z})}}\,,
\end{eqnarray}
where $\rho_0$ stands for the initial probability density subject to the pandemic (or to the diffusion process). In this work, we consider $\rho$ as the probability density of active cases of COVID-19 to make a parallel between our model and real pandemic data. The number of active cases of COVID-19 is defined as follows \cite{world}
\begin{equation}
\mbox{Active Cases} = \mbox{Total Cases} - \mbox{Total Deaths} - \mbox{Recovered}\,,
\end{equation}
therefore, it represents the current number of patients detected and confirmed as infected with SARS-CoV-2. The number of active cases is also a relevant metric for public health and primary care functions, as it allows measuring capacity versus hospitalization needs. 
Such a number is used to plot the distributions of active cases of COVID-19 for different countries as we see in \cite{world}, and it was used to predict the pandemic peaks in different countries as one can see in \cite{tsallis_2020}.
Before starting to make numerical integrals of Eq. (\ref{eq2}) to depict the active cases curves, let us comment on the dimensionality of the model. A (spatial) bi-dimensional model seems most natural to discuss the diffusion process in the population. This usually is captured by the bi-dimensional networks with persons being nodes and relation being links \cite{amati}. Moreover, one particular feature of the transmission of this virus is that it occurs by contact between persons \cite{yang_20}. This feature is captured by a nearest-neighbor interaction between the nodes. It can be argued that one can project down this nearest-neighbor interaction bi-dimensional network into a long-range interaction chain (one-dimensional chain) with appropriate boundary conditions. Thus in the passing to the continuum model we can in a reasonable approximation consider a $D=1$ model with $x$ ranging from $-\infty$ to $+\infty$. For the numerical integration in the reciprocal space, we can take therefore  $-\infty \leq \omega \leq \infty$, and $-\infty \leq k \leq \infty$. Moreover, we choose evaluate $\rho$ at ${\bf x}={\bf x}'$ indicating that the probability density is measured in the same spatial location after the evolution of the diffusion process.

\subsection{Model I - Standard Gaussian solutions}
\label{mod_01}

The simplest model that we can analyze consist in 
\begin{equation}
W_\phi=0\,;\qquad \phi(\tau) = 1\,,
\end{equation}
where the diffusion coefficient is normalized. This first model will enable us to observe the influence of the critical exponent and of the diffusion time in our curves.  

\begin{figure}[h!]
\vspace{1 cm}
 \includegraphics[scale=0.45]{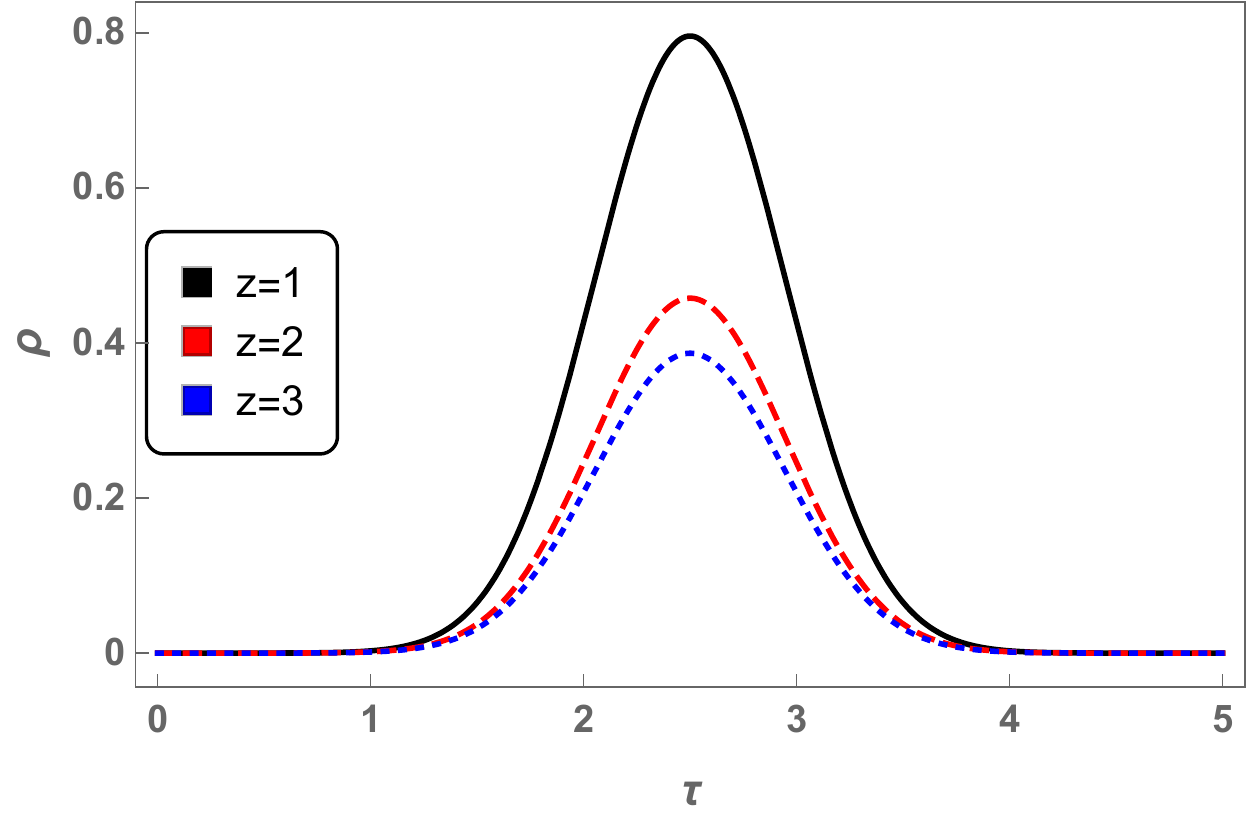} \hspace{1 cm}
 \includegraphics[scale=0.45]{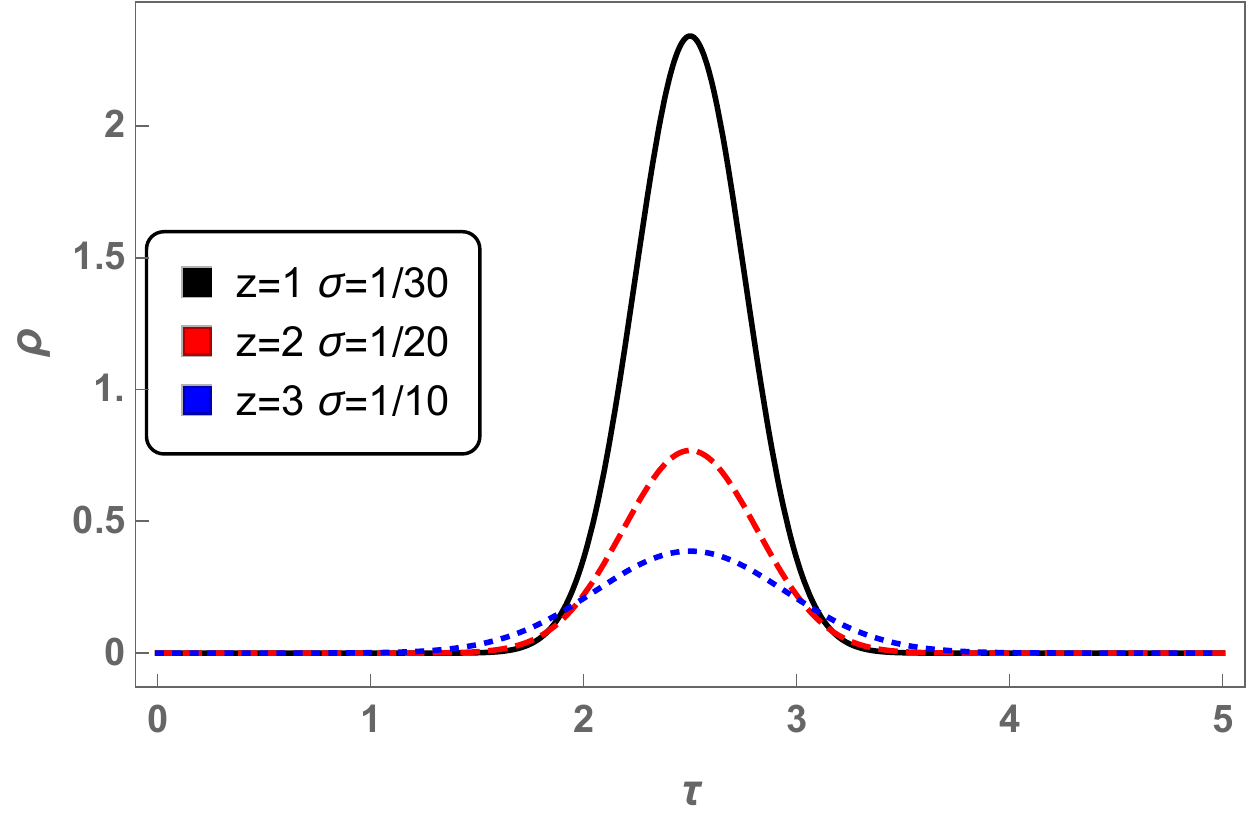}\\
 \includegraphics[scale=0.45]{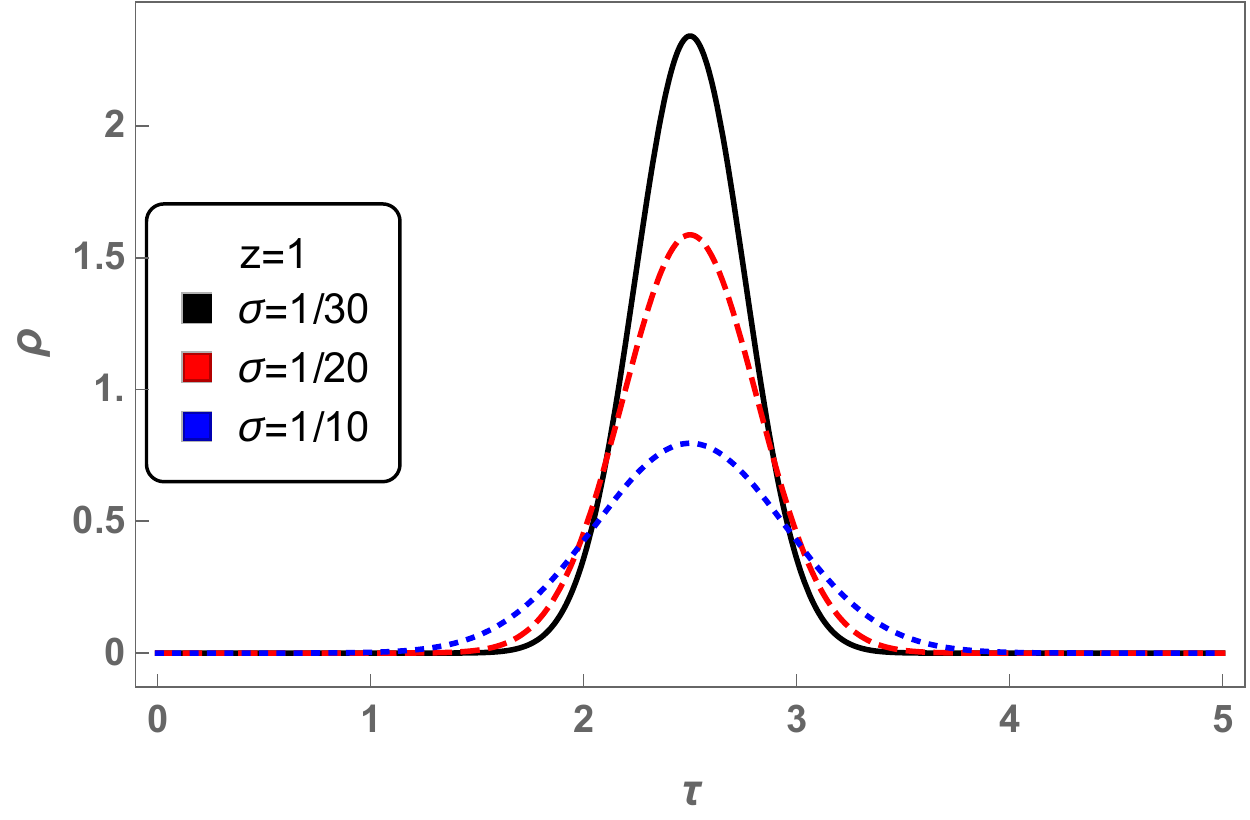} \hspace{1 cm}
 \includegraphics[scale=0.45]{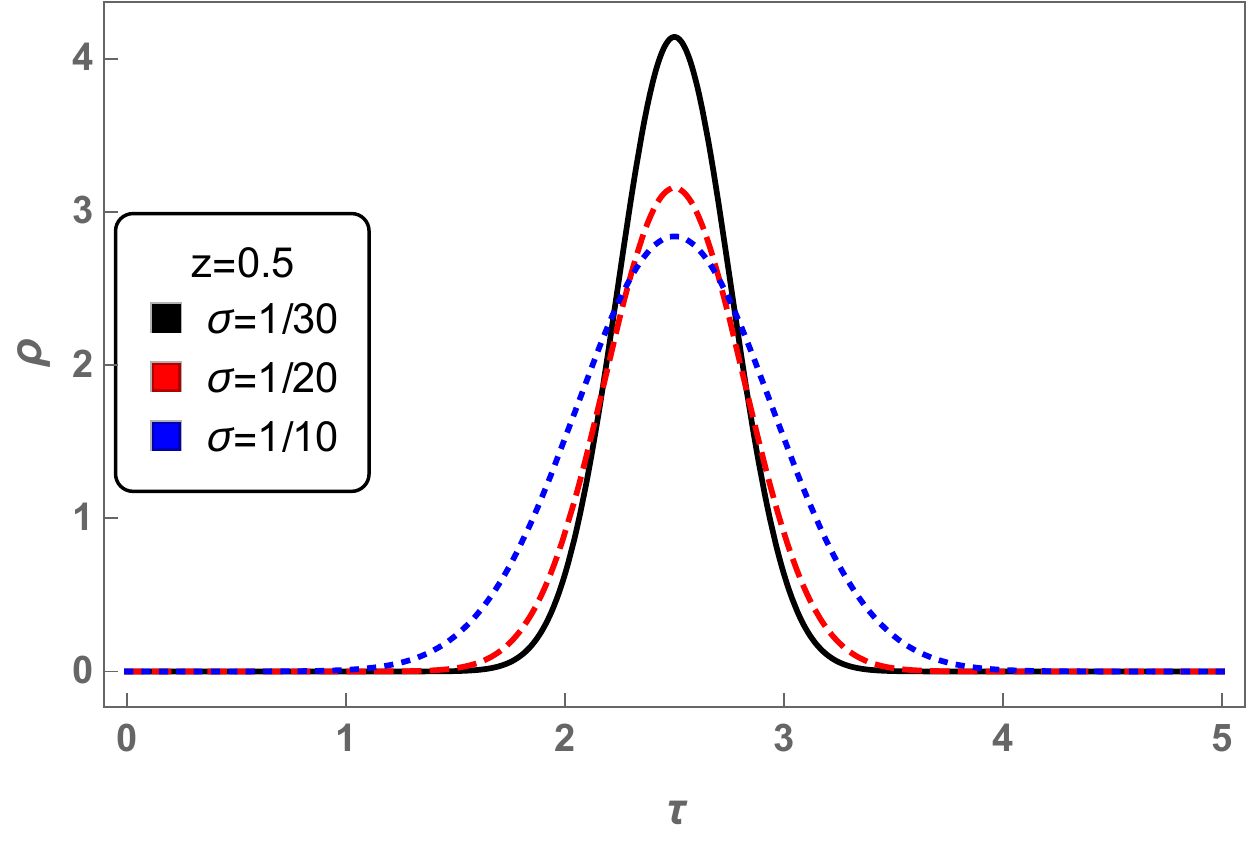}
  \caption{The solution (\ref{eq2}) with $\phi=1$ for $\rho_0=1$, $D=1$, ${\bf x}={\bf x}'$, $\tau'=2.5$ at four scenarios.  {(\it top left}) $\sigma=1/10$, $z=1$ (black solid curve), $z=2$ (red dashed curve), and $z=3$ (blue dotted curve); {(\it top right}) $\sigma=1/30$ and $z=1$ (black solid curve), $\sigma=1/20$ and $z=2$ (red dashed curve), and $\sigma=1/10$ and $z=3$ (blue dotted curve); {(\it bottom left}) $z=1$, $\sigma =1/30$ (black solid curve), $\sigma=1/20$ (red dashed curve), $\sigma=1/10$ (blue dotted curve); and {(\it bottom right}) $z=1/2$, $\sigma=1/30$ (black solid curve), $\sigma=1/20$ (red dashed curve), and $\sigma=1/10$ (blue dotted curve). }
   \label{fig:1}
\end{figure}

The Fig.~\ref{fig:1} shows four different behaviors for the solution of the diffusion equation. The diffusion time and the critical exponent $z$ are competing parameters. Particularly in the top left curve, with a same diffusion time large values of $z\geq1$ makes the Gaussian flattened, whereas bottom right for $z<1$ the Gaussian tends to blow up even for sufficiently large diffusion time. The best scenario in the sense of flattening the Gaussian curve seems to be the increasing of both diffusion time and critical exponent $z$ (top right). The features of $\sigma$ presented in the previous scenarios yield us to indeed understand the diffusion time as equivalent to a lockdown period.

In this work we adopted the lockdown definition introduced by Flaxman et al. \cite{imp_rep_0320}, which means a scenario where regulations and laws regard strict social interaction. These regulations/laws include the banning of any non-essential public gatherings, closure of educational, public and cultural institutions, and ordering people to stay at home apart from exercise or essential tasks.

\subsection{Model II - Solutions for a (anti)kink-like diffusion coefficient}
\label{mod_02}

\begin{figure}[h!]
\vspace{1 cm}
 \includegraphics[scale=0.45]{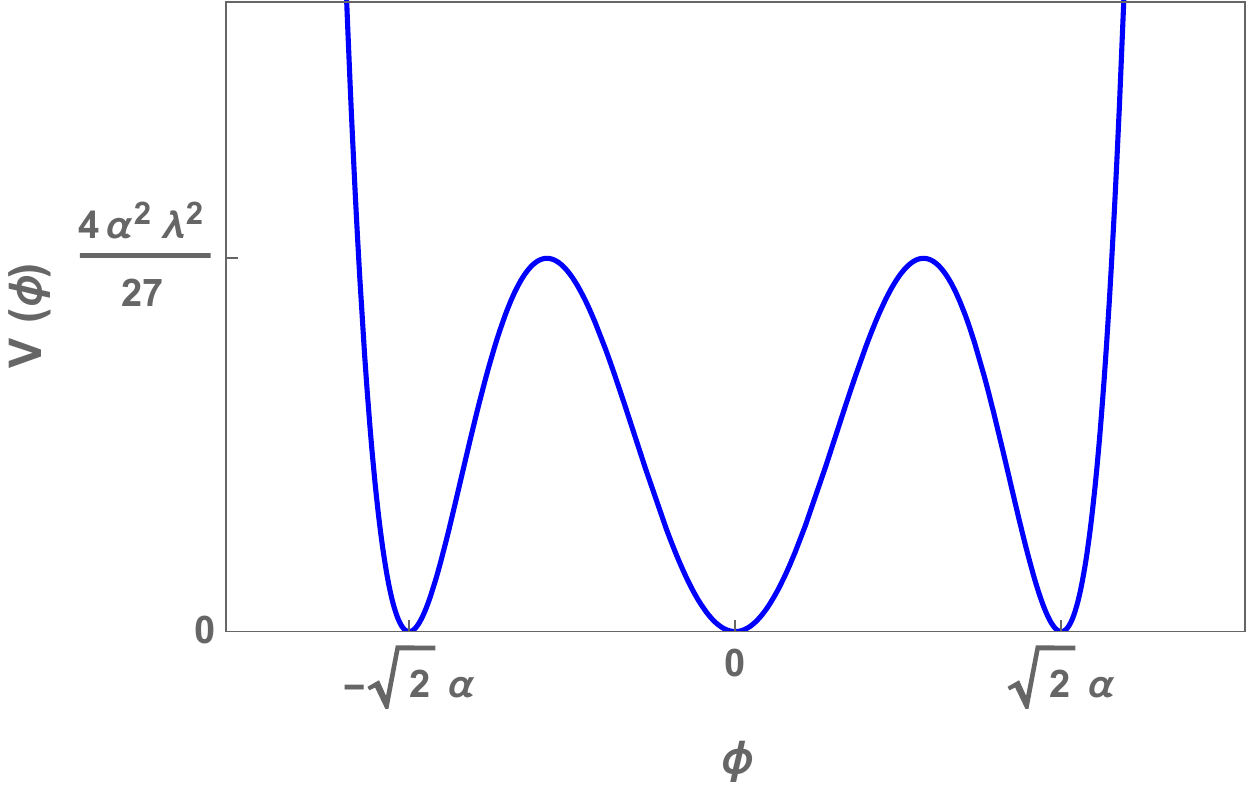}\hspace{1 cm}
 \includegraphics[scale=0.45]{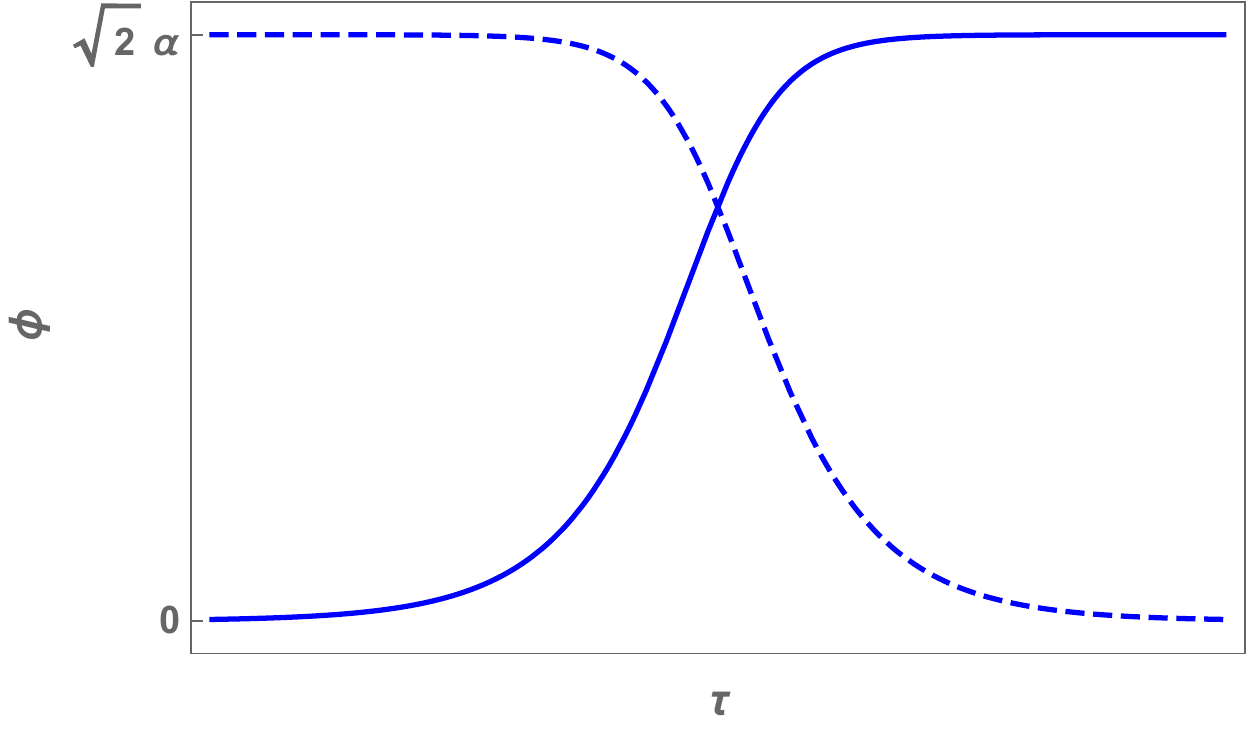}
  \caption{In the left panel we depicted potential $V(\phi)$ Eq. (\ref{eq0201}), and in the right panel we present solution $\phi$ Eq. (\ref{eq0202}), for $s=1$ yielding to a kink (blue solid curve), and for $s=-1$ leading to an anti-kink (blue dotted curve).}
   \label{fig:2_0}
\end{figure}

In this example we are going to consider the following superpotential
\begin{equation}
W=\frac{\lambda}{8}\,\phi ^2\, \left(4-\frac{\phi ^2}{\alpha ^2}\right)\,,
\end{equation}
where $\lambda$ and $\alpha$ are real constants. Such superpotential leads us to 
\begin{equation} \label{eq0201}
V=\frac{\lambda^2}{4}\,   \phi^2  \left(2-\frac{\phi ^2}{\alpha ^2}\right)^2\,.
\end{equation}
which is known as $\lambda\,\phi^6$ potential, and it is depicted in the left panel of Fig. \ref{fig:2_0}. The minima or vacua values of $V$ correspond to $\phi_v=0$, and $\phi_v=\pm\sqrt{2\,\alpha}$. The region between these different minima is called a topological sector, and it mediates the transmission rate in our pandemic model. This potential was applied in subjects like higher-order phase transitions in Ginzburg–Landau theory \cite{saxena,almeida}, modeling domain walls in ferroelastic transitions \cite{hatch}, and in new physics beyond the standard model of particles at high energies scales \cite{LISA, zhang}. The previous definition of $W$ yields to the first-order differential equation
\begin{equation}
\phi_{\,\tau}=\pm\,\frac{\lambda}{2}\,   \phi\,  \left(2-\frac{\phi ^2}{\alpha ^2}\right)\,,
\end{equation}
whose analytic solutions are
\begin{equation} \label{eq0202}
\phi(\tau)=\alpha\,  \sqrt{1+s\,\tanh (\lambda  \,(\tau -\tau_1))}\,;\qquad s=\pm 1\,,
\end{equation}
exhibiting anti-kink and kink-like profiles as one can see in Figure \ref{fig:2_0}. Once $\phi(\tau)$ can be interpreted as the transmission rate of the virus, in Fig. \ref{fig:2_0}, we observe that for $\tau\ll \tau_1$ the virus is not transmitted, however its transmission increases until an approximately constant rate ($\phi_v=\sqrt{2\,\alpha}$) as $\tau$ gets bigger than $\tau_1$. The applicability of such a model is going to be carefully discussed later, in section \ref{act_cases}.

\subsection{Model III - Solutions for a double (anti)kink-like diffusion coefficient}
\label{mod_03}

\begin{figure}[h!]
 \includegraphics[scale=0.45]{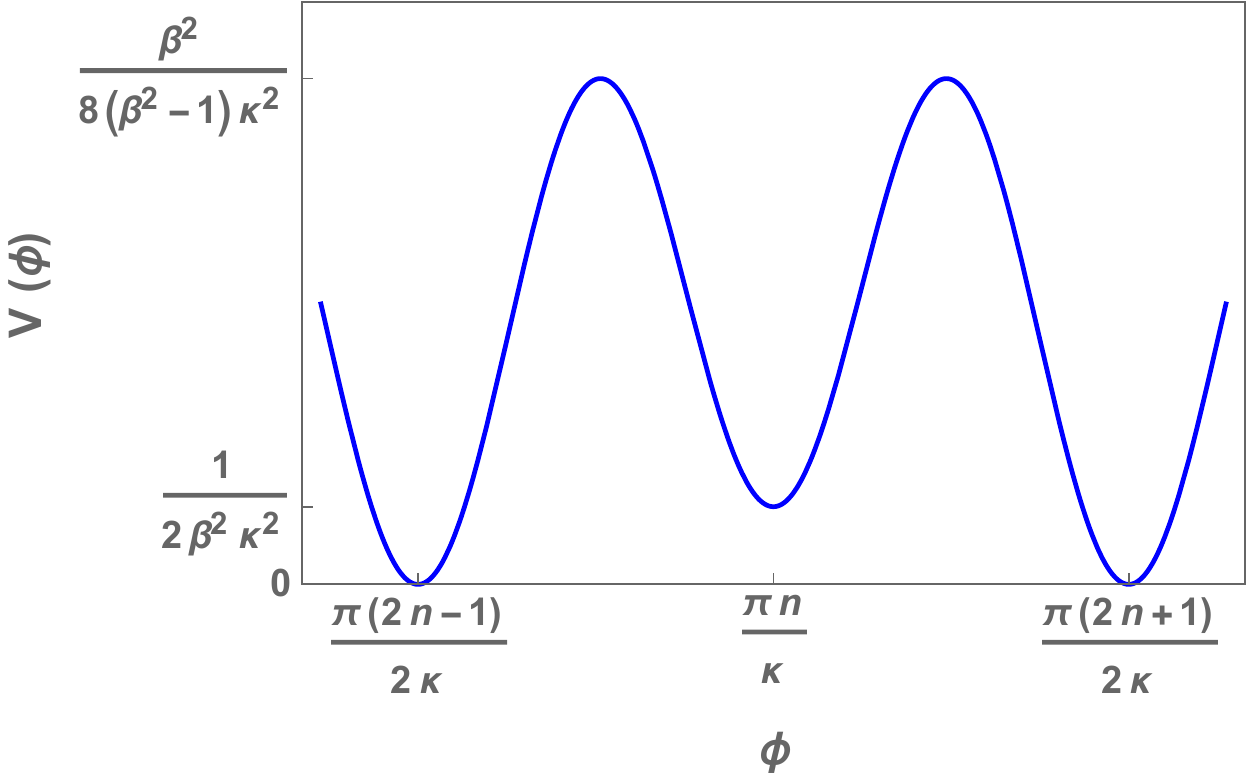}\hspace{1 cm}
 \includegraphics[scale=0.45]{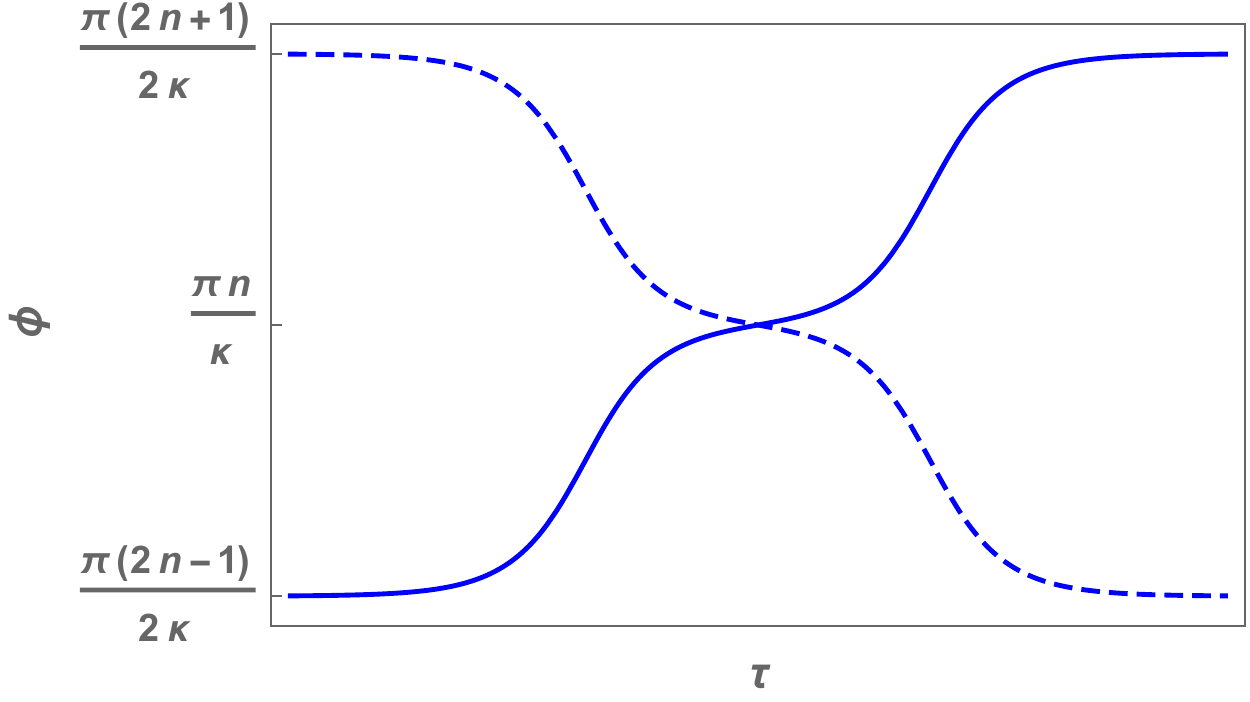}
  \caption{In the left panel we depicted potential $V(\phi)$ Eq. (\ref{eq0301}), and in the right panel we present solution $\phi$ Eq. (\ref{eq0302}), for $s=1$ yielding to a double kink (blue solid curve), and for $s=-1$ leading to a double anti-kink (blue dotted curve).}
   \label{fig:3_0}
\end{figure}

As a next example, let us deal with a more complex model for the diffusion coefficient. Such a model is derived from
\begin{equation} \label{eq0301}
V=\frac{\cos ^4(\kappa \, \phi ) \,\left(\beta ^2 \tan ^2(\kappa \, \phi )+1\right)}{2\,\beta\, \kappa }\,,
\end{equation}
this potential is known as double sine-Gordon model and it is applied to study ultra short optical pulses in He$^{3}$ \cite{bull, muss}, in decaying of false vacuum and phase transitions in field theory \cite{muss, samira}. 
The left panel of Fig. \ref{fig:3_0} shows part of the periodic form of potential $V$ (\ref{eq0301}). There we can see one topological sector between vacua $\phi_v=\pm\,n\,\pi/(2\,\kappa)$. An interesting feature about this model is that the parameter $\beta$ can deform this potential at $\phi=2\,\pi\,n/\kappa$. Such deformation is responsible for the double (anti)kink profiles observed in the right panel of Fig. \ref{fig:3_0}, and it is also related to the variation of the transmission rate of our pandemic model. 

The previous potential yields to the first-order differential equation
\begin{equation}
\phi_\tau=\pm\frac{\cos ^2(\kappa  \phi ) \sqrt{\beta ^2 \tan ^2(\kappa  \phi )+1}}{\beta  \kappa }\,,
\end{equation}
which is satisfied by the analytic solutions
\begin{equation} \label{eq0302}
\phi(\tau)=\frac{1}{\kappa}\,\left(s\tan^{-1}\left(\frac{\sinh (\tau -\tau_1)}{\beta }\right)+n\,\pi\right)\,,
\end{equation}
for $s=\pm 1$, and $n=0,\pm1,\pm2,...\,\,$. The behavior of these analytic solutions $V$ can be appreciated in Fig. There it is shown that $\phi(\tau)$ has three different regimes of transmission rate, the first one for $\tau \ll \tau_1$ where $\phi_v=\,(2\,n-1)\,\pi/(2\,\kappa)$,  the second regime occurs when $\tau \approx \tau_1$ and $\phi\approx \pi\,n/\kappa$, and the third one appears for $\tau \gg \tau_1$ and $\phi_v=\,(2\,n+1)\,\pi/(2\,\kappa)$. These regimes can lead us to distributions with different waves of contagious, as we are going to show below. In the next section we are going to analyze the viability of model III and compare the numerical curves of $\rho$ derived from it with those obtained through model II.

\section{Modeling active cases of COVID-19}
\label{act_cases}

As it is known, the active cases data can present a high level of uncertainty once it depends on the number of the tests performed by each country and also the countries' transparency in reporting the tests. 
The data set used to depict the graphics of this section were taken up to May 02, 2020. Therefore, some discrepancies between our predictions and the pandemic evolution are expected. To depict active cases curves that could reproduce the existent data and able to predict the behavior of the pandemic, we decided first to test our model against a now well-established data set, and for that, we choose data from Germany. Nowadays Germany is the 6th leading country in the world in numbers of COVID-19 cases, accumulating a total of $164967$ cases up to May 02 \cite{world_g}. Moreover, Germany is the third leading country in the total number of tests, reporting a total amount of $2547052$ tests for SARS-CoV-2, which corresponds to $30400$ tests per million of population (considering the data up to May 02) \cite{world_g}. Therefore, to constraint some free parameters in our model, we use the data from Germany as guidance. Despite this procedure, we still have other free parameters to represent the particular features of each country's strategy to deal with the pandemic. We present the features of our models against real active cases data from Germany, Brazil, and Spain below.

\subsection*{Germany}

\begin{figure}[h!]
\vspace{1 cm}
 \includegraphics[scale=0.45]{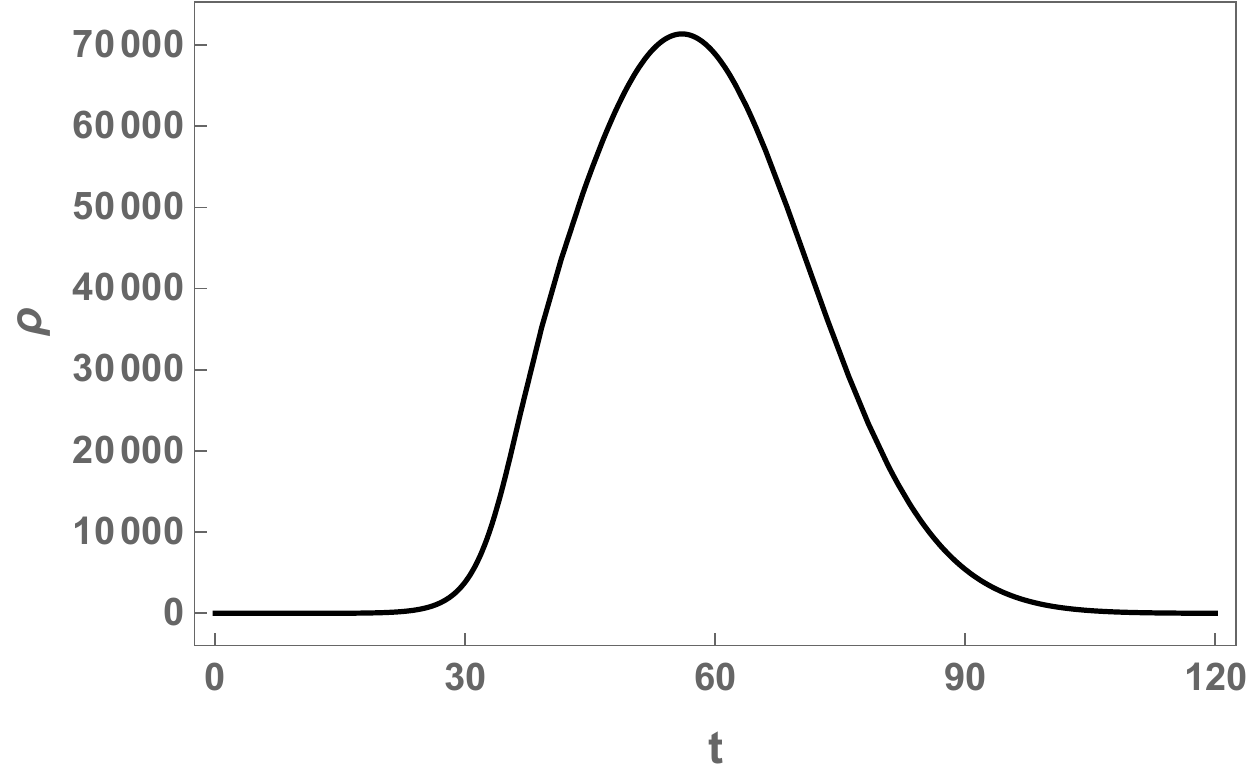}\hspace{1 cm}
 \includegraphics[scale=0.45]{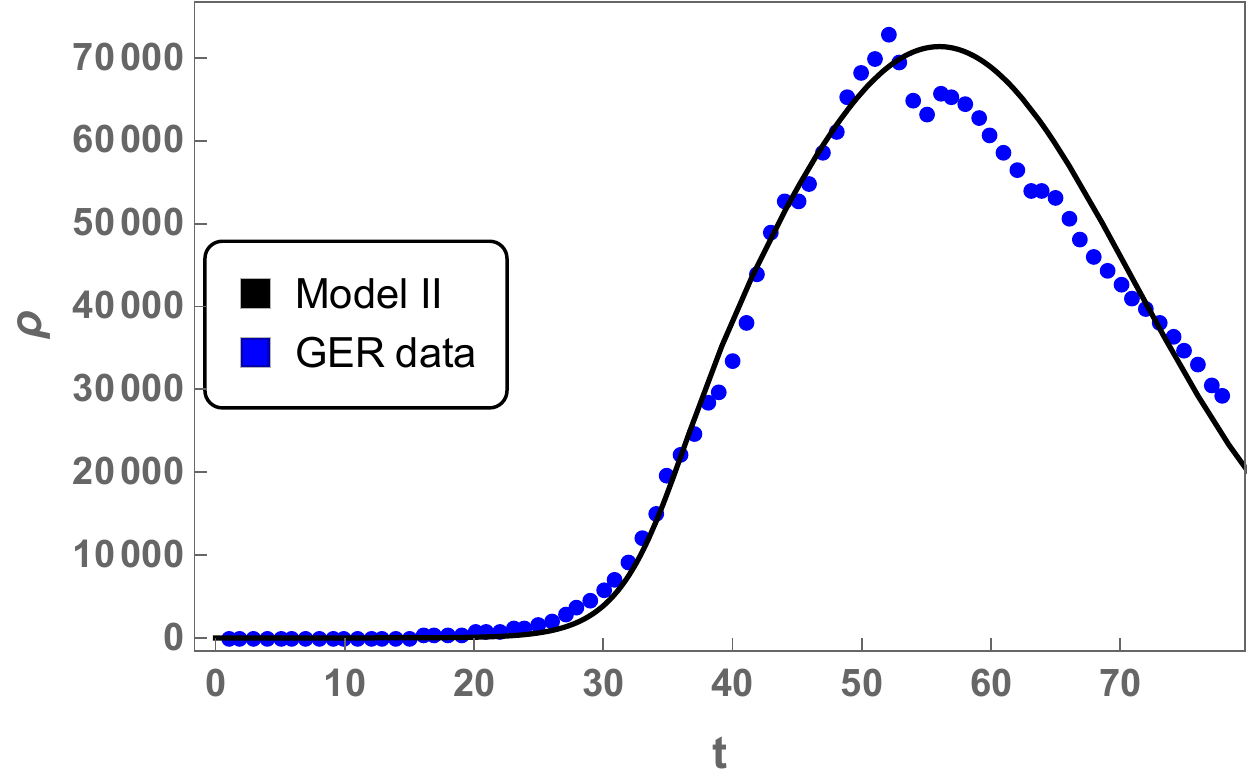}
  \caption{In these graphics we present the active case solution for $t=2 \tau$ (time in days), $D=1$, ${\bf x}={\bf x}'$, $\rho_0=10^{7}$, $\tau^{\,\prime}=28$, $\tau_1=17.6$, $\alpha =0.65$, $s=1$, $\lambda=0.5$, $z=2.5$, and $\sigma=28$. We also depicted the active cases data from Germany (blue dots), since February 15, 2020 (day 1), until May 02, 2020 (day 78) \cite{world_g}.}
   \label{fig:2_1}
\end{figure}

\begin{figure}[h!]
 \includegraphics[scale=0.45]{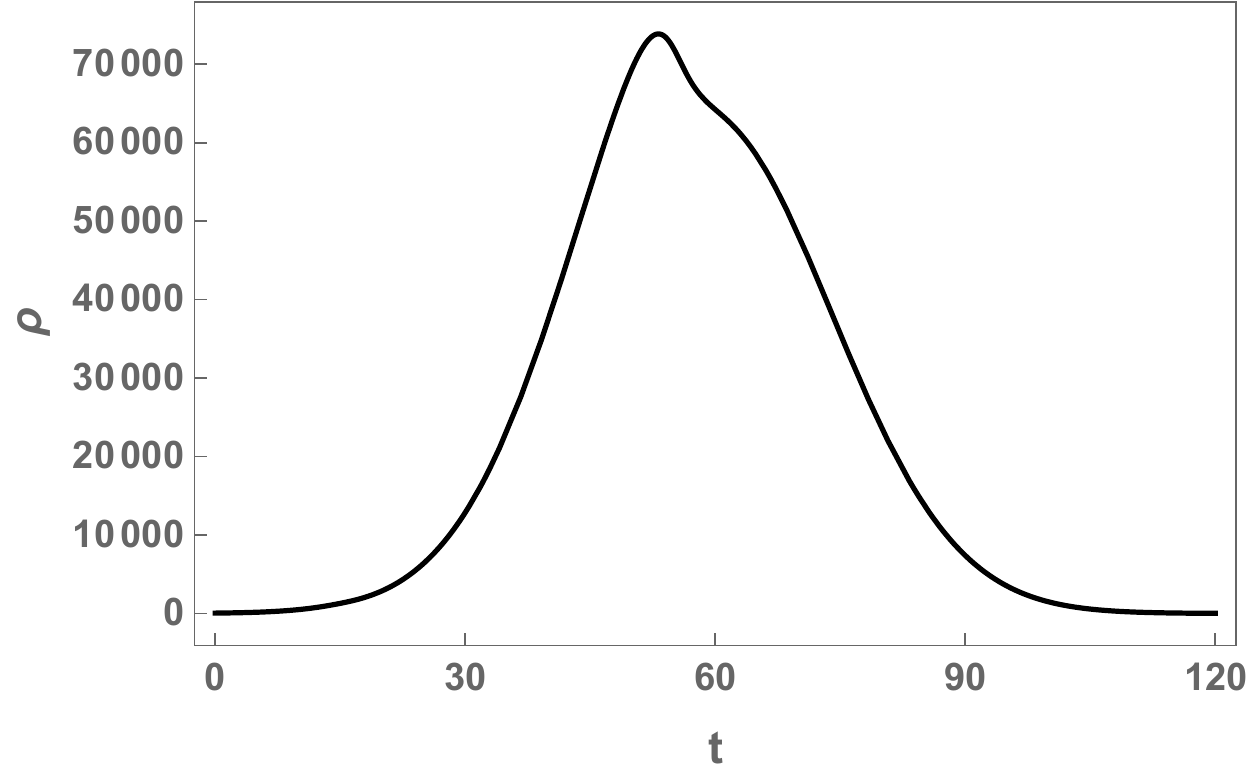} \hspace{1 cm}
 \includegraphics[scale=0.45]{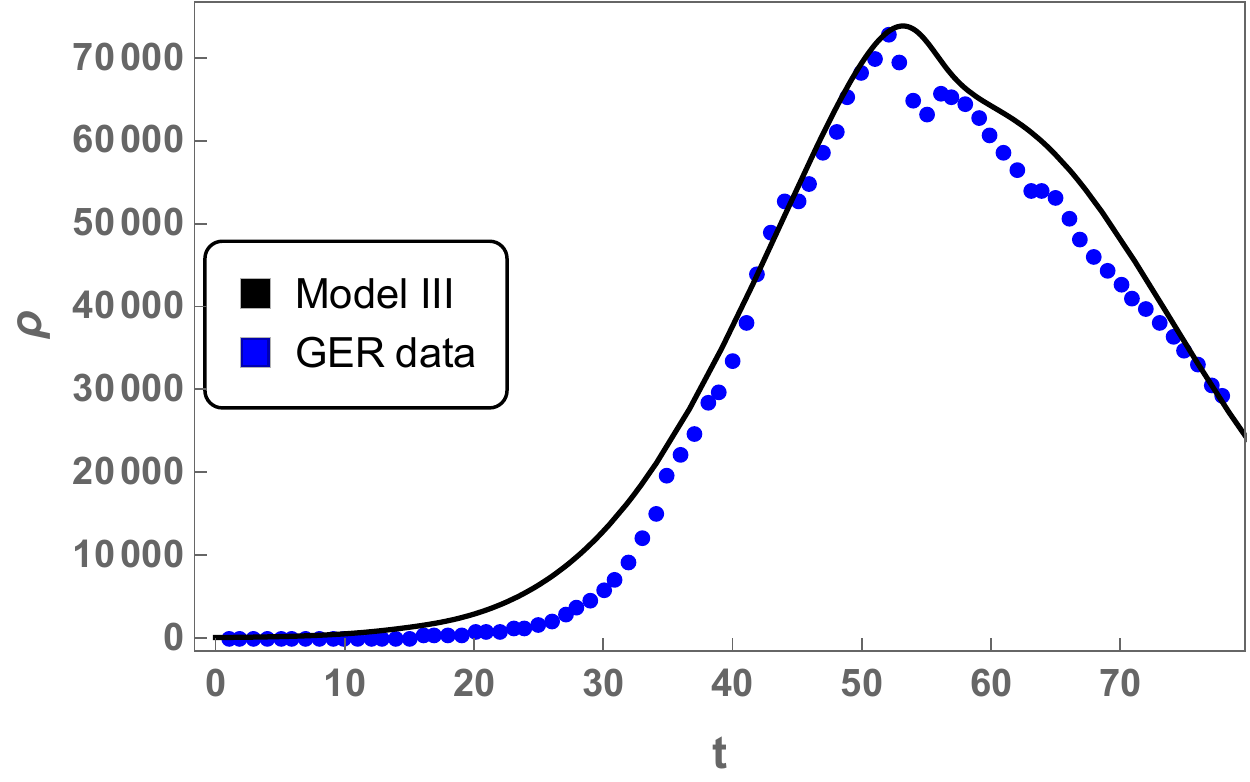}
  \caption{In these graphics we present the active case solution for $t=2 \tau$ (time in days), $D=1$, ${\bf x}={\bf x}'$, $\rho_0=10^{5}$, $\tau^{\,\prime}=29.5$, $\tau_1=17.75$, $\kappa =0.058$, $s=-1$, $\beta=10^{\,4}$, $z=2.5$, $\sigma=28$, and $n=2$. We also depicted the active cases data from Germany (blue dots), since February 15, 2020 (day 1), until May 02, 2020 (day 78) \cite{world_g}.}
   \label{fig:3_1}
\end{figure}

In this first scenario we shown in Figs. \ref{fig:2_1}, and \ref{fig:3_1}, the active cases curves for Germany integrated from model II and III, respectively. There, the black solid curves are the numerical solutions generated from Eq.(\ref{eq2}), which best fitted the real data depicted in blue dots. The parameters constrained in the fitting process were $\rho_0 = 10^{7}$ (for Fig \ref{fig:2_1}), $\rho_0=10^{5}$ (for Fig. \ref{fig:3_1}), and time scaling $t=2\,\tau$ representing the number of days of pandemic. Moreover, we also used $\sigma = 28$ (for Figs. \ref{fig:2_1}, and \ref{fig:3_1}), referring to the number of lockdown days in Germany \cite{lockdown}. 

The left panels of Figs. \ref{fig:2_1}, and \ref{fig:3_1} predict that the pandemic of COVID-19 would be fully controlled in Germany after day $102$ of infection (or after May 26, 2020), when the number of active cases is less than $1000$ people.  Besides, the active cases curves unveil that the kink and the double anti-kink solutions, Eqs. (\ref{eq0202}), and (\ref{eq0302}) deform the standard Gaussian curve. By comparing the right panels of Figs. \ref{fig:2_1}, and \ref{fig:3_1}, we realize that the double anti-kink solution yields us to a better fitting of real data, reproducing a change in the decreasing of the number of active cases at $t\approx 60$ day.

\subsection*{Spain}

\begin{figure}[h!]
\vspace{1 cm}
 \includegraphics[scale=0.45]{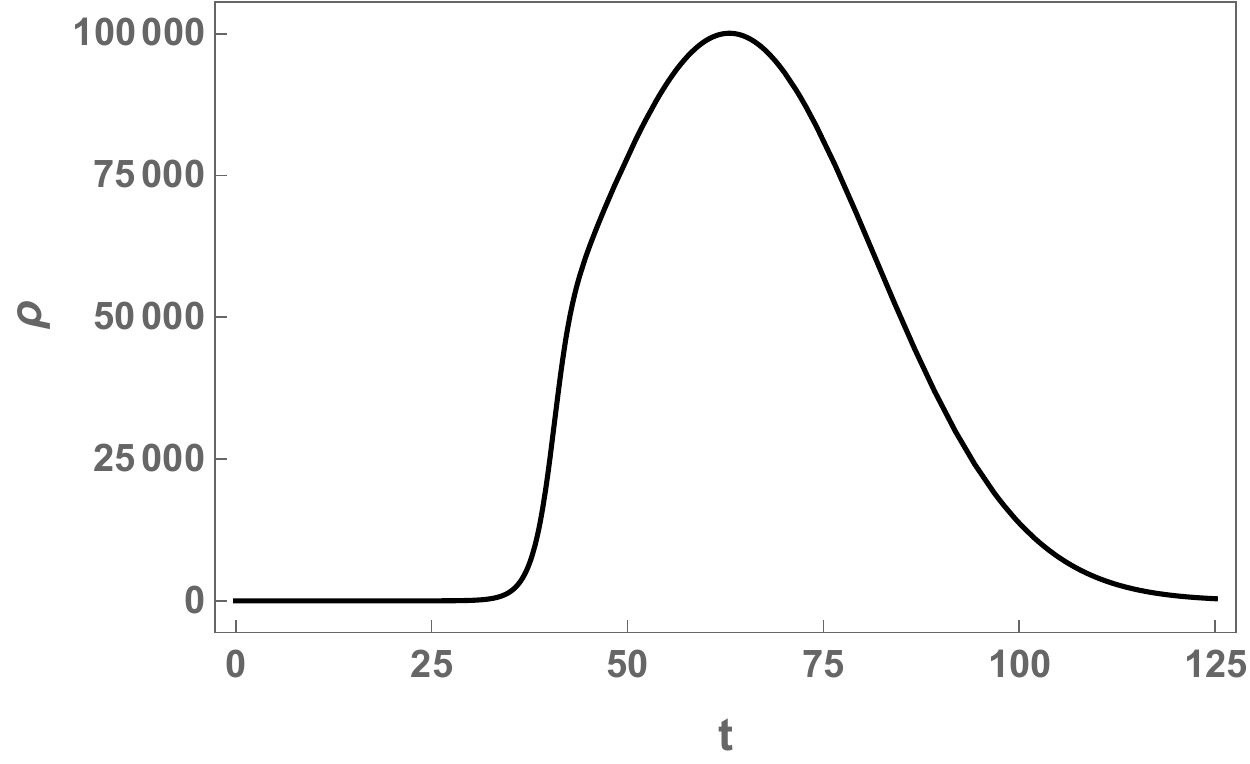}\hspace{1 cm}
 \includegraphics[scale=0.45]{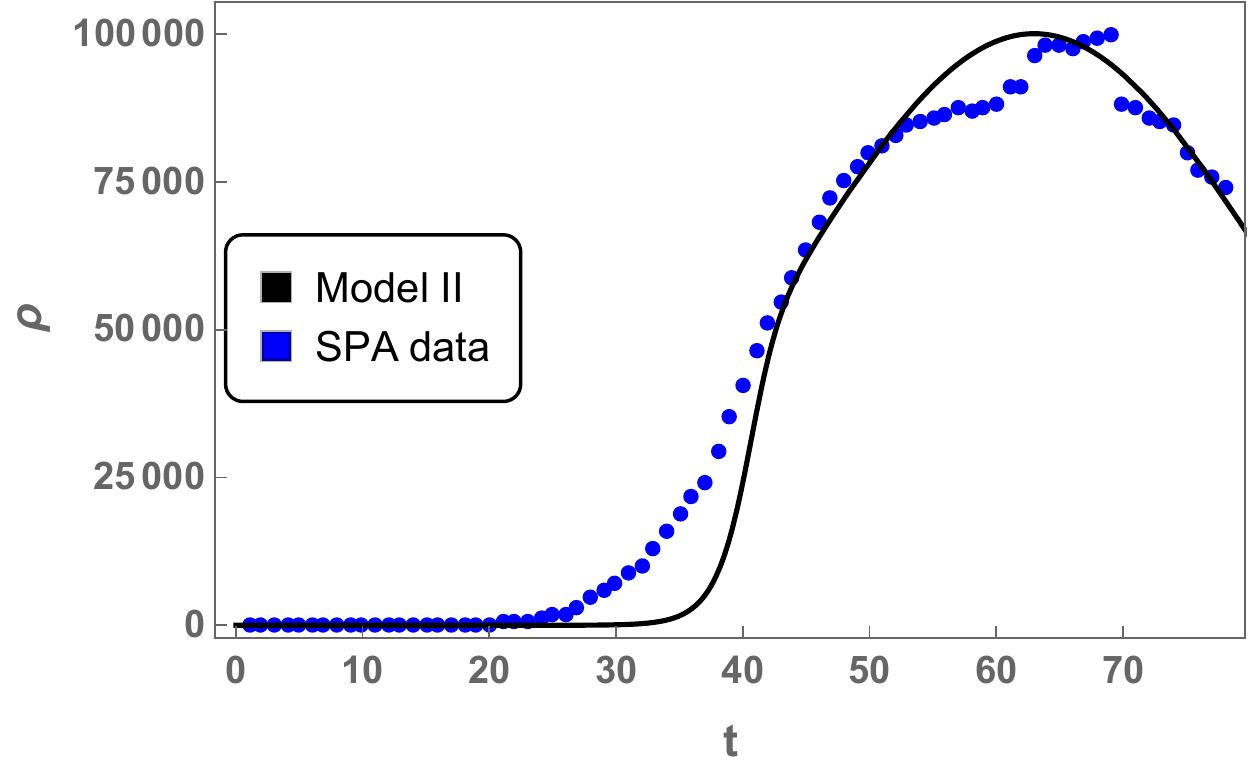}
  \caption{In these graphics we present the active case solution for $t=2 \tau$ (time in days), $D=1$, ${\bf x}={\bf x}'$, $\rho_0=10^{7}$, $\tau^{\,\prime}=31.5$, $\tau_1=20.5$, $\alpha =0.27$, $s=1$, $\lambda=1$, $z=50$, and $\sigma=43$. We also depicted the active cases data from Spain (blue dots), since February 15, 2020 (day 1), until May 02, 2020 (day 78) \cite{world_s}.}
   \label{fig:2_2}
\end{figure}

\begin{figure}[h!]
\vspace{1 cm}
 \includegraphics[scale=0.45]{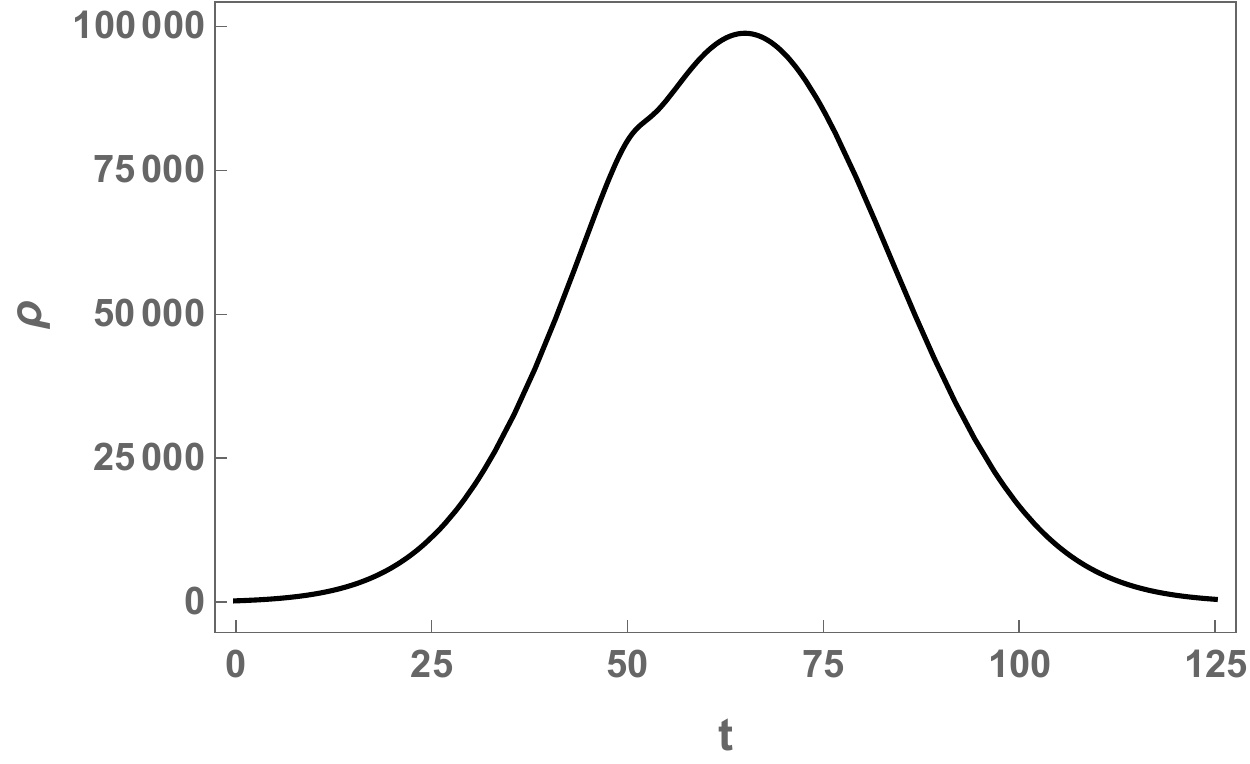}\hspace{1 cm}
 \includegraphics[scale=0.45]{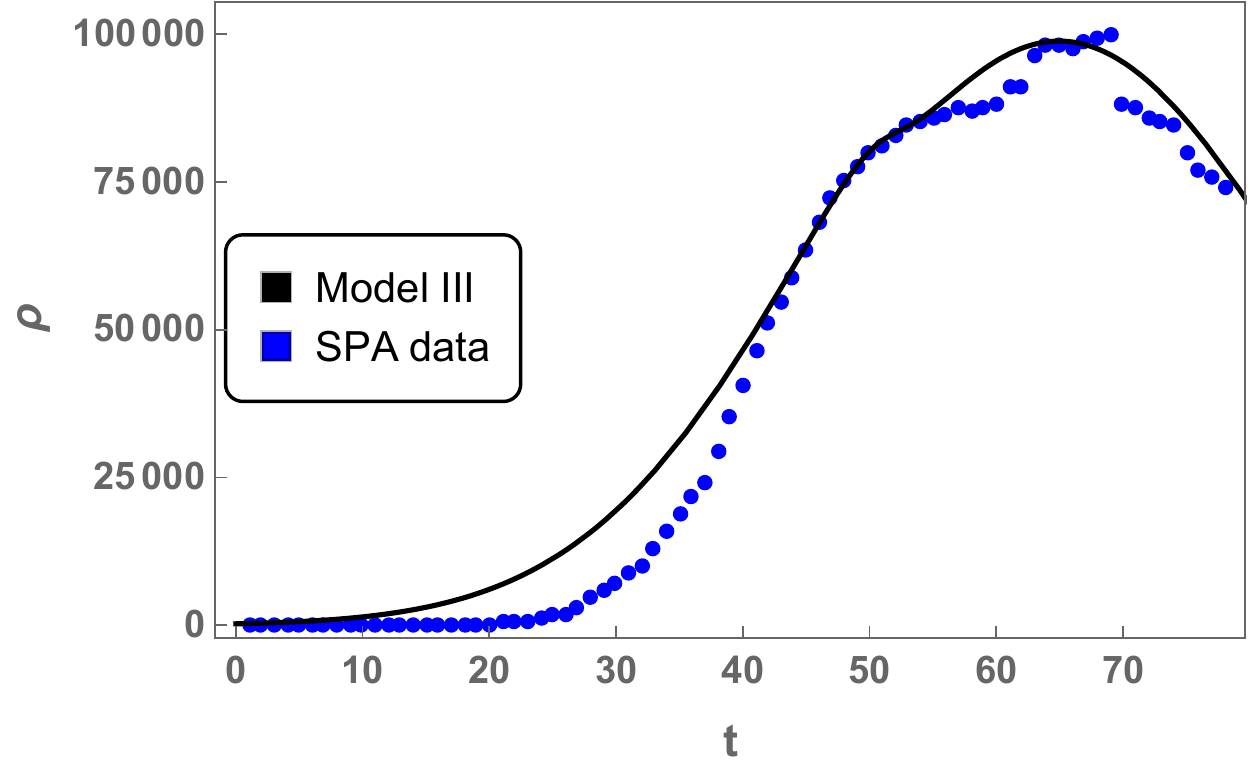}
  \caption{In these graphics we present the active case solution for $t=2 \tau$ (time in days), $D=1$, ${\bf x}={\bf x}'$, $\rho_0=10^{5}$, $\tau^{\,\prime}=32.5$, $\tau_1=45$, $\kappa =0.25$, $s=-1$, $\beta=10^{8}$, $z=50$, $\sigma=43$, and $n=3$. We also depicted the active cases data from Spain (blue dots), since February 15, 2020 (day 1), until May 02, 2020 (day 78) \cite{world_s}.}
   \label{fig:3_2}
\end{figure}

The graphics of Figs. \ref{fig:2_2}, and \ref{fig:3_2} shown the currently infected people by SARS-CoV-2 in Spain. At the present moment, Spain reported an amount of $245567$ cases of COVID-19, which makes it the second leading country in the world pandemic rank \cite{world_s}. The active cases data for Spain are depicted as blue dots in Figs. \ref{fig:2_2}, and \ref{fig:3_2}, and to plot our numerical curves we considered $\sigma=43$ lockdown days \cite{lockdown}.  We can observe that the black solid curves are in good agreement with the pandemic data, and they predict that the active cases of COVID-19 would be fully controlled in Spain after day $120$ (June 11, 2020), where the number of infected people is less than $1000$. Moreover, in this scenario the kink and the double anti-kink like solutions Eqs. (\ref{eq0202}), and (\ref{eq0302}), strongly deform the standard Gaussian curve. As in the case of Germany, the double anti-kink curve leads us to a better fitting of real data, reproducing a second wave of contagious after day $50$.

\pagebreak

\subsection*{Brazil}

\begin{figure}[h!]
\vspace{1 cm}
 \includegraphics[scale=0.45]{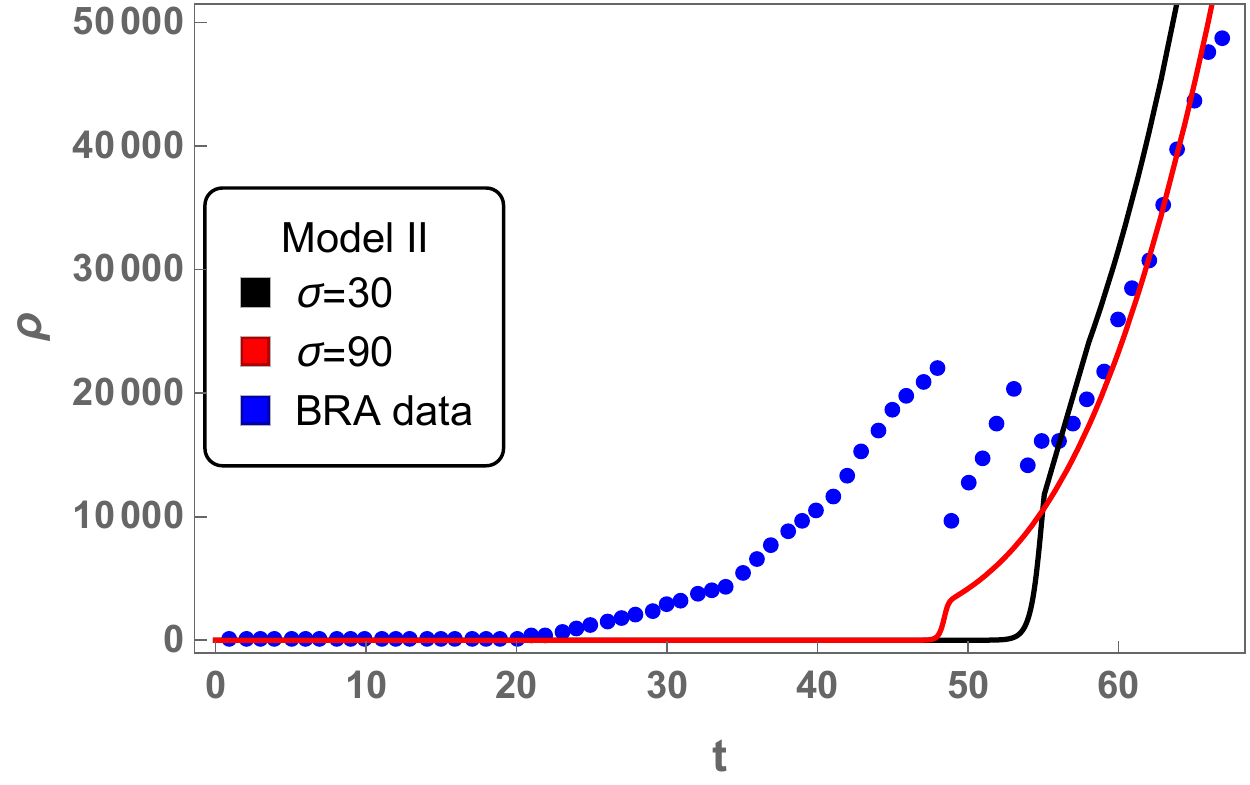}\hspace{1 cm}
 \includegraphics[scale=0.45]{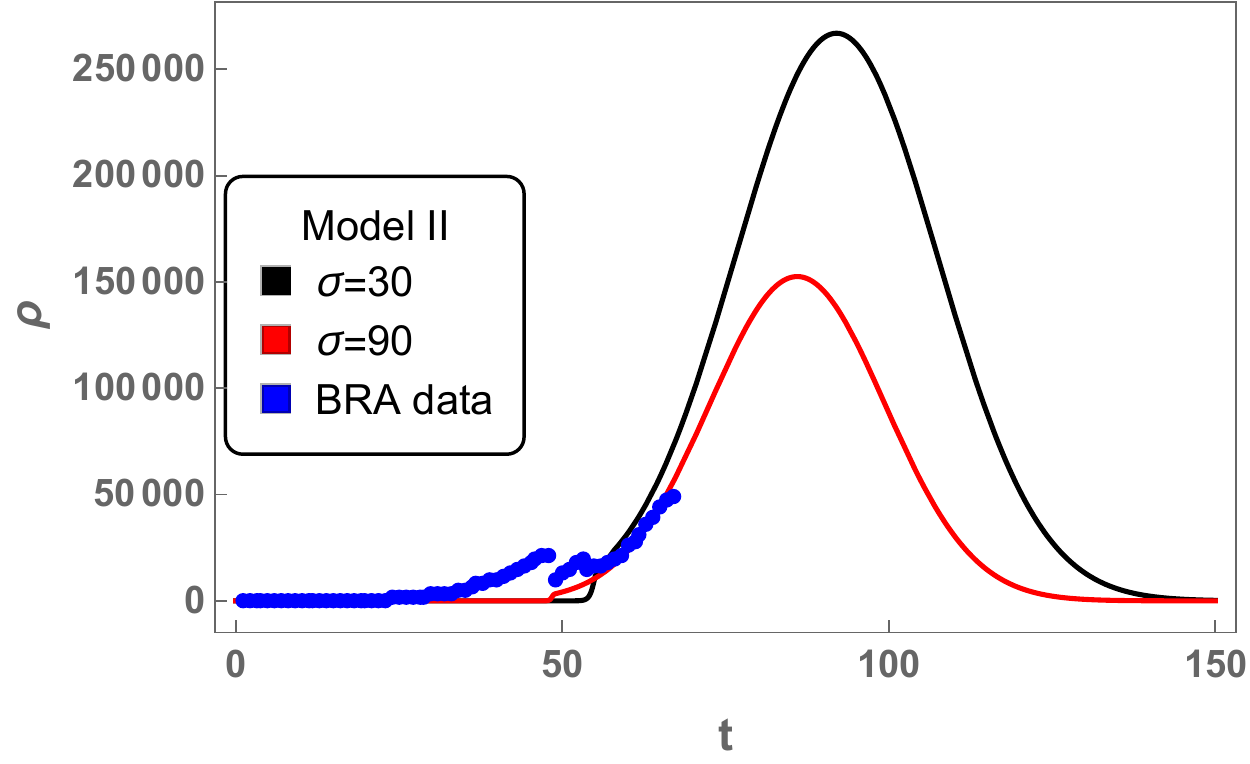}
  \caption{In these graphics we present the active cases solutions for $t=2 \tau$ (time in days), $D=1$, ${\bf x}={\bf x}'$, $\rho_0=10^{7}$, $\alpha =0.6$, $s=1$, $\lambda=4$, $z=50$, $\tau^{\,\prime}=42$, $\tau_1=27.5$, and $\sigma=30$ (black solid curve), besides $\tau^{\,\prime}=86$, $\tau_1=48.5$, and $\sigma=90$ (red solid curve). The peaks of black and red curves are approximately $268000$ and $153000$ cases, respectively. We also depicted the active cases data from Brazil (blue dots), since February 26 (day 1), 2020 until May 02, 2020 (day 67) \cite{world_b}.}
   \label{fig:2_3}
\end{figure}

\begin{figure}[h!]
\vspace{1 cm}
 \includegraphics[scale=0.45]{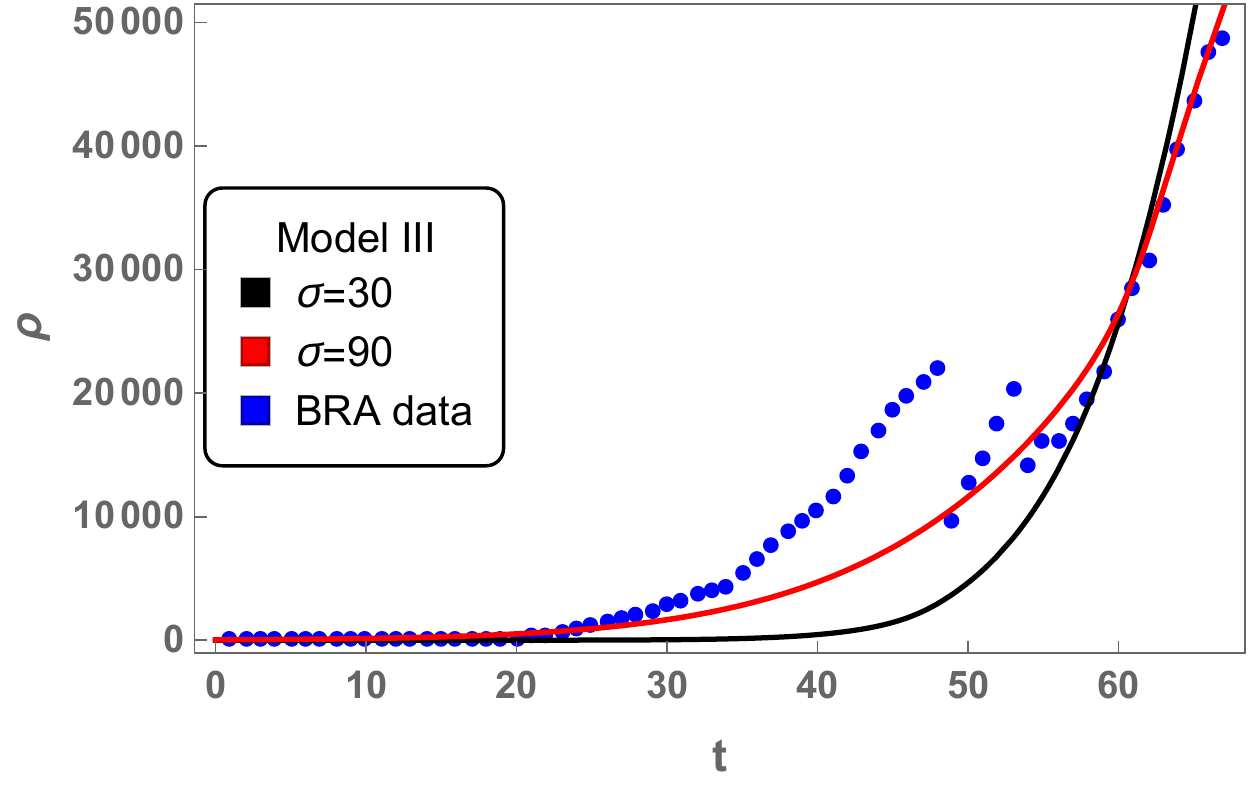} \hspace{1 cm}
 \includegraphics[scale=0.45]{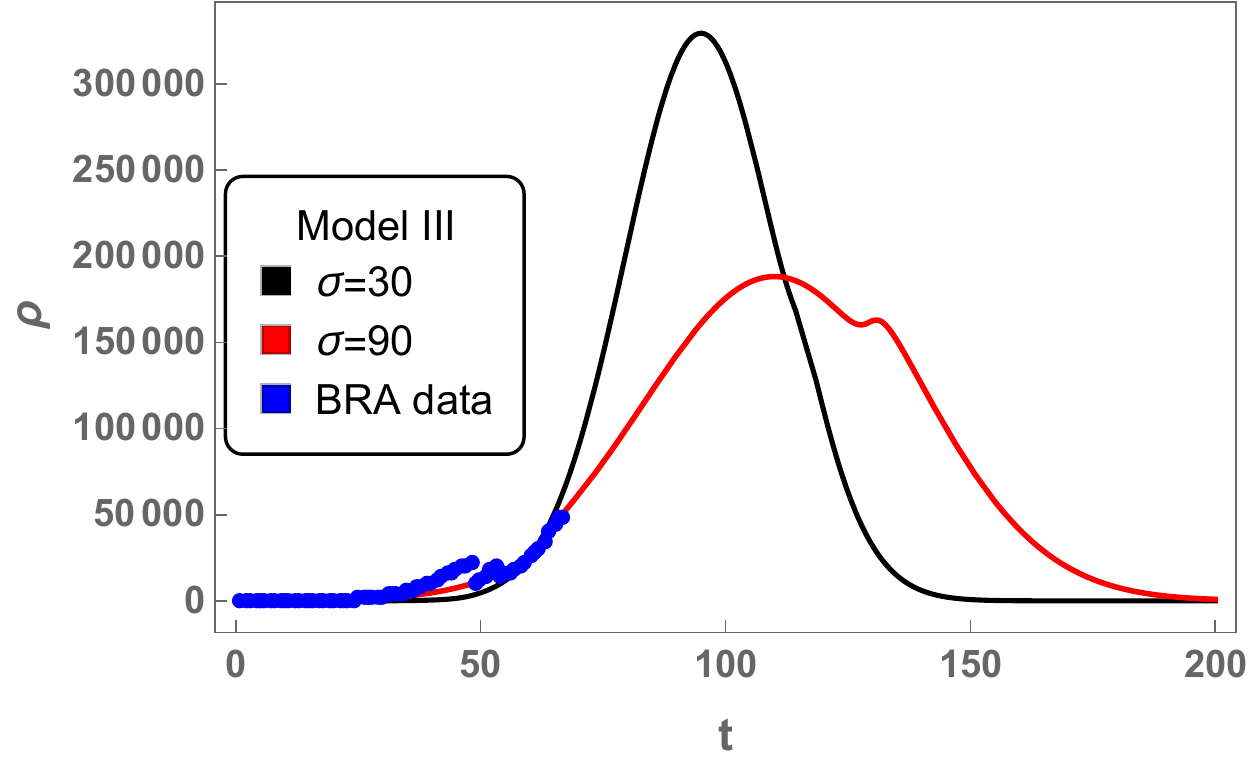}
  \caption{In these graphics we present the active cases solutions for $t=2 \tau$ (time in days), $D=1$, ${\bf x}={\bf x}'$, $\rho_0=10^{5}$,  $\kappa =0.06$, $s=1$, $\beta=10^{\,7}$, $n=2$, $z=50$, $\tau^{\,\prime}=47.5$, $\tau_1=40.5$, $\sigma=30$ (black solid curve), besides $\tau^{\,\prime}=55$, $\tau_1=48$, and $\sigma=90$ (red solid curve). The peaks of black and red curves are approximately $329000$ and $188000$ cases, respectively. We also depicted the active cases data from Brazil (blue dots), since February 26, 2020 until May 02, 2020 \cite{world_b}.}
   \label{fig:3_3}
\end{figure}

The pandemic of COVID-19 starts in Brazil eleven days after spreads in Europe. Despite this few difference in time, the evolution of the contamination in Brazil was deeply different from Germany and Spain, as we can see in the graphics of Figs. \ref{fig:2_3}, and \ref{fig:3_3}. The numerical solutions presented in Figs. \ref{fig:2_3}, and \ref{fig:3_3} were depicted using the same values for $\rho_0$ and $t$ from Germany curves and the same value for $z$ from Spain curves. Moreover, we derived two possible scenarios of lockdown measures, the black solid curves show a model with $\sigma=30$ lockdown days while the red solid curves were depicted with $\sigma=90$ lockdown days. 

Furthermore, we also included the active cases data from Brazil as blue dots \cite{world_b}. The real data from Brazil reveal an abrupt change of contamination which started on day 48 (April 13, 2020) and developed to a new increasing rate after day 56 (April 26, 2020). Along this period, besides the pandemic, Brazil has been facing a political crisis, which would explain the behavior of the COVID-19 infection here observed. Another problem with Brazil's data is its high degree of uncertainty. Up to May 02, 2020, Brazil reported a total of $97100$ cases of COVID-19, figuring at the 9th leading country in the world pandemic rank \cite{world}. However, Brazil has performed only $339552$ tests of COVID-19, representing a total of $1597$ per million of people \cite{world}. 

For these reasons, it is challenging to make any prediction about the evolution of the pandemic in Brazil. Although, our curves seem to be in good agreement with the active cases data so far, and we also can realize that a long time social distance flatten the curves of the active cases. The peaks of the two active cases curves from Figs. \ref{fig:2_3}, and \ref{fig:3_3} have considerable differences about $115000$, and $141000$ cases, respectively. Moreover, the peaks are predicted to happen on days $92$ (May 27, 2020), and $86$ (May 21, 2020) for black and red solid curves from Fig. \ref{fig:2_3}, and on days $95$ (May 30, 2020), and $110$ (June 04, 2020) for black and red curves from Fig. \ref{fig:3_3}. It is relevant to mention, that up to May 02, 2020, Brazil can attend in maximum $32703$ patients with needs for ICU \cite{leitos}. The solutions from Fig. $\ref{fig:2_3}$ predict that the pandemic in Brazil would not be fully controlled earlier than day $131$ (June 22, 2020). In the longer predicted scenario, observed in the red solid curve of Fig. \ref{fig:3_3}, the pandemic would be fully controlled after day $198$ (August 28, 2020), when the number of active cases is less than $1000$. We also realize that the double anti-kink like solutions are able to fit better the real data than the single kink-like ones, reproducing the increasing in the number of active cases after day $48$.

\section{Modeling coupled spread of COVID-19}
\label{coupled}

As it is known the primary mechanism used by COVID-19 to spread is through person-to-person contact. Consequently, big cities favor the spreading of the virus to small centers, once they facilitate the mixing of people from different areas. Moreover, their strategical positions close to airports and crossed by state roads, make such cities susceptible to rapidly spread the virus to innermost regions. This phenomenon of advancing of COVID-19 into the countryside happened in several places in the world, such as around New York City and close to several metropolises from Brazil. It was also modeled in several regions of Brazil using a Markov chain approach as one can see in the work of Costa et al. \cite{costa_2020}.  In this section, we adapted our diffusion equation to describe the dissemination of the virus around different cities. In order to reproduce such a scenario, we rewrite our diffusion equation as

\begin{equation} \label{eq0401}
\frac{\partial}{\partial\sigma}\rho_i({\bf x},\tau;{\bf x}',\tau';\sigma)=\,\phi(\tau)\left(\frac{\partial^2}{\partial\tau^2} +(-1)^{z+1}\Delta^z\right)\phi(\tau)^{\,-1}\,\rho_i({\bf x},\tau;{\bf x}',\tau';\sigma)+J_i\left(\sigma\right)\,\phi(\tau)\,; \qquad i= \,1,\,2,\, \mbox{...},\, N\,,
\end{equation}
where $N$ is the number of the cities, and $J_i(\sigma)$ are free sources of the diffusion process. The previous equations present the following general analytic solutions

\begin{eqnarray} \label{eq0402}
\rho_i({\bf x},\tau;{\bf x}',\tau';\sigma)=\rho_{\,0i}\,\int{\frac{d\omega\, d^D{\bf k}}{(2\pi)^{D+1}}\,\phi(\tau)\,e^{i\omega(\tau-\tau')+i{\bf k}\cdot({\bf x}-{\bf x}')}\,e^{-\sigma(\omega^2+|{\bf k}|^{2z})}}+\int^{\sigma}\,d\bar{\sigma}\,J_i\left(\bar{\sigma}\right)\,\phi(\tau)\,,
\end{eqnarray}
and the free sources should obey the conservation constraint
\begin{equation}
\sum_{i=1}^{N}\,J_i(\sigma) = 0\,.
\end{equation}
The simplest definition for the sources is to work with $J_i(\sigma)=\mbox{constant}$. In this application we are going to understand such sources as proportional to the basic reproductive rate at the beginning of the pandemic, commonly known as $R_0$ \cite{anderson}. The basic reproductive rate is a non-dimensional quantity which measures the secondary cases of contamination produced by one case introduced in susceptible populations \cite{yang_01}. Such a rate is a key ingredient in several mathematical models to describe pandemic scenarios, and any attempt to estimate its value is a real challenge.
Consequently, by working with an equivalence between $J_i$ and $R_0$, our model suggests that the basic reproductive rate is changed between different cities as the pandemic evolves in a given region.

To exemplify our methodology, let us consider two cities from Brazil which experienced the phenomenon of the spreading of SARS-COV-2 to innermost regions. The cities here considered are S\~ao Paulo, and S\~ao Jos\'e dos Campos. These two cities have about $12$ million and $721$ thousand of inhabitants, respectively, and they are approximately $100\,km$ distant apart. They are experiencing the so-called yellow-phase of the reopening plan designed by the state of S\~ao Paulo government, where people have access to public places such as parks, restaurants, and cultural events with limited capacity \cite{plano_sao_paulo}.

The pandemic started in S\~ao Paulo at February 25 (day 1), and up to August 02, 2020 it accumulated more than $200000$ cases with more than $9600$ fatalities \cite{data_sp}. The first case of COVID-19 in S\~ao Jos\'e dos Campos was reported in March 18, 2020 and the city has been registered more than $17000$ cases, besides $204$ deaths up to August 02, 2020 \cite{data_sp}. Moreover, the most recent basic reproductive rate estimated for the state of S\~ao Paulo in the beginning of the pandemic is $R_0 = 9.24$ \cite{yang_02}. In such an application, the best model to fit the cumulative cases data from the two cities was Model $II$, therefore, the asymptotic behavior of $\rho$ in Eq. ($\ref{eq0402}$) is going to be proportional to the product $J_{\,i}\,\phi_v$, corresponding to a kink like profile for the density distribution. So, in the following application, we choose to test our model against the cumulative cases data.  

\begin{figure}[h!]
\vspace{1 cm}
 \includegraphics[scale=0.45]{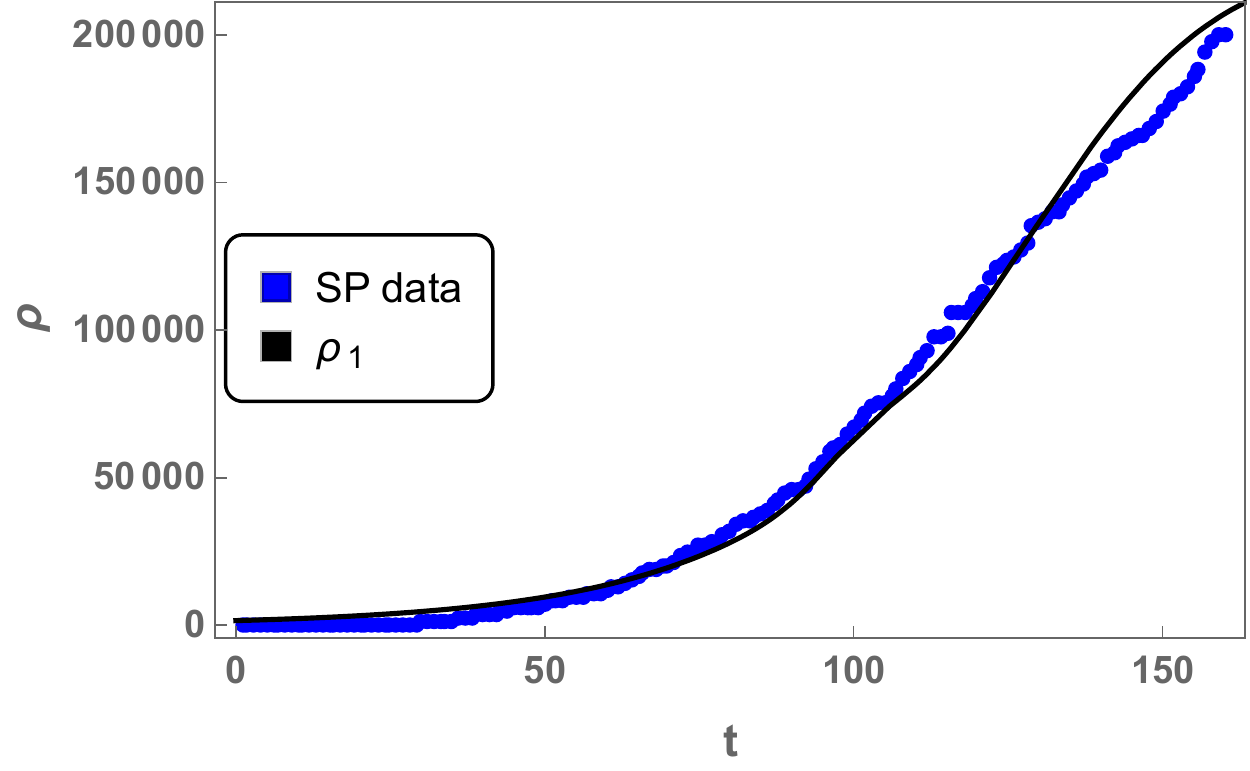}\hspace{1 cm}
 \includegraphics[scale=0.45]{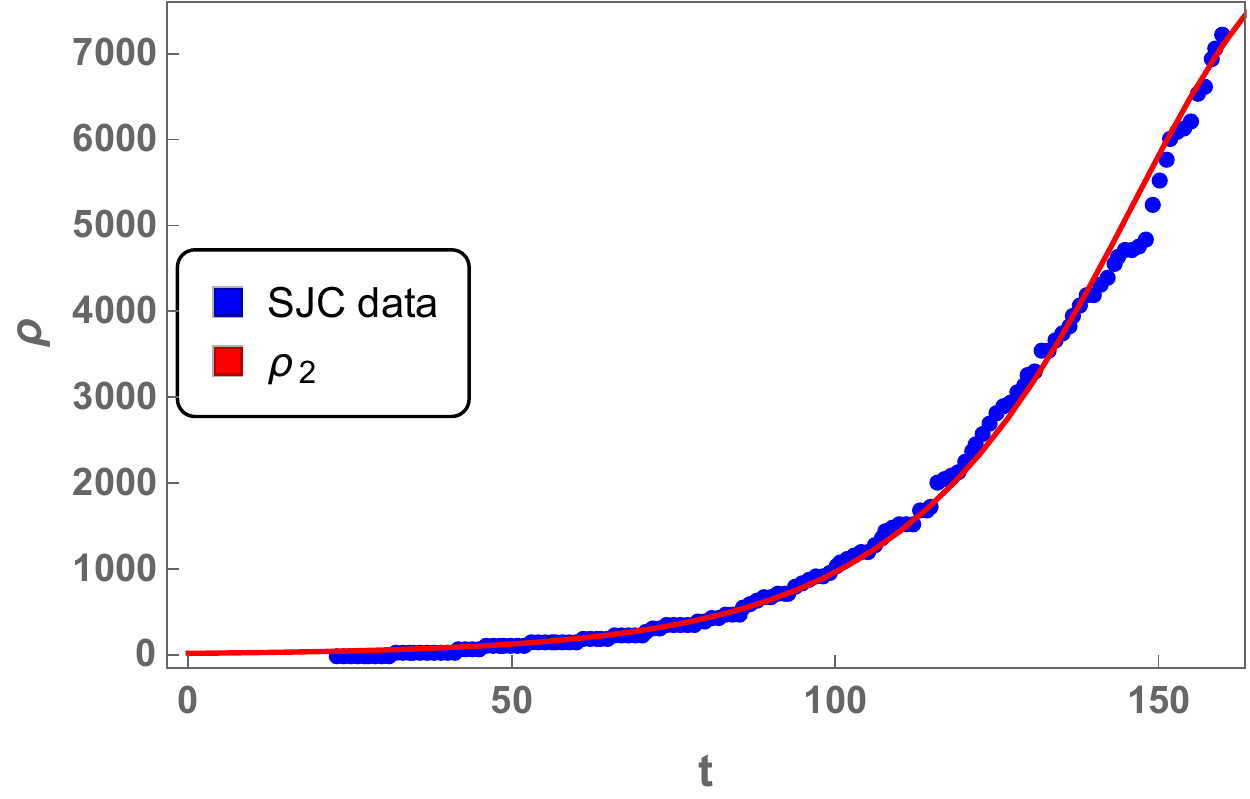}
  \caption{In the left panel we present the cumulative cases solutions for Model II with $t=2 \tau$ (time in days), $D=1$, ${\bf x}={\bf x}'$, $\rho_0=10^{2}$, $\alpha =2.90\times 10^{3}$, $s=1$, $\lambda=0.72\times 10^{-1}$, $z=50$, $\tau^{\,\prime}=50$, $\tau_1=69$, $J_{\,1} = 9.24 $, $\tilde{\rho}_0=1$, and $\sigma=6$ (black solid curve).  We also depicted the cumulative cases data from the city of S\~ao Paulo (blue dots), since February 25 (day 1), 2020 until August 02, 2020 (day 160)\cite{data_sp}. In the right panel we present the cumulative cases solutions for Model II with $t=2 \tau$ (time in days), $D=1$, ${\bf x}={\bf x}'$, $\rho_0=10$, $\alpha =-117.38$, $s=1$, $\lambda=0.82\times 10^{-1}$, $z=50$, $\tau^{\,\prime}=85$, $\tau_1=77.5$, $J_{\,2} = -9.24 $, $\tilde{\rho}_0=1$, and $\sigma=6$ (red solid curve). Besides, we depicted the cumulative cases data from the city of S\~ao Jos\'e dos Campos (blue dots), since March 18 (day 1), 2020 until August 02, 2020 (day 137) \cite{data_sp}.}
   \label{fig:4_1}
\end{figure}

Then, let us consider $N=2$ as the number of the cities, and 
\begin{equation}
J_1=-J_2=\tilde{\rho}_0\,R_0\,,
\end{equation}
where $\tilde{\rho}_0$ is a proportionality constant and $R_0$ is the basic reproductive rate. Taking the previous ingredients into Eq. (\ref{eq0402}) yield us to depict the graphics presented in Fig. \ref{fig:4_1}. There, we consider labels $i=1$, and $i=2$ to describe S\~ao Paulo and S\~ao Jos\'e dos Campos, respectively, and we worked with $\sigma = 6$ lockdown days, corresponding to the anticipation of holidays in the city of S\~ao Paulo \cite{feriados}. Such anticipation of holidays was planned to increase the social distancing, attempting to reduce the spreading of COVID-19. We can observe that our model successfully reproduces the evolution of the pandemic in theses two cities if we consider  $z=50$, which is the same value used to fit Spain and Brazil's active cases curves in the previous section. Moreover, the fact that $J_2=-R_0$ means that the basic reproduction rate was passing from S\~ao Paulo to S\~ao Jos\'e dos Campos between February 25 and March 18, as the pandemic spreading evolves.

\section{Conclusions}
\label{conclusions}

In this work we introduced a modified version of the diffusion equation, mediated by a Lifshitz scaling together with the diffusion coefficient $\phi(\tau)$. The diffusion time $\sigma$ is analogous to the so-called fictitious time in stochastic quantization theories \cite{namiki}. We were able to find an analytic solution for this diffusion equation, and to use the standard Gaussian curves to interpret $\sigma$ as the lockdown time, if such an equation is applied to model pandemic cases.

Therefore, we investigate two possible models with $\phi(\tau)$ having (anti)kink, and double (anti)kink-like profiles. These models were used to fit real active cases data of COVID-19 from three different countries (Germany, Spain, and Brazil).  We successfully depicted the active cases curves for Germany and Spain up to May 02, 2020, and use them to constraint some of our free parameters to generate curves for Brazil. Then, we predicted four scenarios for the advance of the pandemic in Brazil, based on $30$ and $90$ lockdown days. These scenarios alerted for a potential escalation of the pandemic in  Brazil if no lockdown measure is taken. Moreover, the solution which best fitted the active cases data up to May 02, 2020, is the red curve from Fig. \ref{fig:3_3}, where it was considered $90$ lockdown days. Such a solution predicted that the pandemic would be fully controlled after day $198$ (August 28, 2020). From the previous analyses we are able to observe how crucial the lockdown measures are to flatten the active cases curves and to control the pandemic spread, corroborating with the remarks from \cite{canabarro_2020}.

We also applied our model in a new phase of the pandemic, where the virus is moving towards innermost regions of the countries. In this application we considered several diffusion processes interacting through free sources. Each one of these diffusion processes corresponds to the spreading of SARS-COV-2 in a given city. As a simplest case to model, we worked with constant sources and understood them as proportional to the so-called basic reproductive rate at the beginning of the pandemic ($R_0$). In order to exemplify our model we built the cumulative cases curves of two cities from Brazil - S\~ao Paulo and S\~ao Jos\'e dos Campos, and compare them with real data. Our results unveil that this pandemic model can be successfully applied in the context of coupled spreading of COVID-19.

It is relevant to point that the Lifshitz scaling exponent was essential to depict the numerical curves here studied. Moreover, we also verified that the double (anti)kink-like profiles were able to fit better the real active cases data than the single (anti)kink-like solutions. Consequently, we can conjecture that multiple anti(kink)-like models, like those introduced in \cite{dutra}, would improve the level of precision of our active cases curves, and would enable us to model multiple pandemic phases. Another interesting perspective consists in investigate solutions for the diffusion equation derived from models with multiple lump like solutions, such as those presented in \cite{new_models}. Moreover, the methodology here adopted can be applied to other pandemics, as well as to other COVID-19 data set like new daily cases, for instance.

\acknowledgments

We would like to thank CNPq, CAPES and PRONEX/CNPq \& Paraiba State Research Foundation (Grants no. 165/2018 and 0015/2019), for partial financial support. MAA, FAB, EP and JRLS acknowledge support from CNPq (Grant nos. 306962/2018-7, 312104/2018-9, 304852/2017-1 and 420479/2018-0, respectively). The authors also would like to thank the anonymous referees for thoughtful comments which undoubtedly raised the quality of this work.

\end{document}